\documentclass[useAMS,usenatbib]{mn2e}
\usepackage[dvips]{graphicx}
\usepackage{amssymb}

\def\aj{AJ}%
\def\araa{ARA\&A}%
\def\apj{ApJ}%
\def\apjl{ApJ}%
\def\aap{A\&A}%
\def\aaps{A\&AS}%
\def\mnras{MNRAS}%
\def\nat{Nature}%
\title[Halos around edge-on disk galaxies in the SDSS]{Halos around edge-on disk galaxies in the SDSS}
\author[S. Zibetti et al.]{Stefano Zibetti$^{1}$\thanks{E-mail:
zibetti@MPA-Garching.MPG.DE}, Simon D. M. White$^{1}$ and Jon Brinkmann$^{2}$\\
$^{1}$Max-Planck-Institut f\"ur Astrophysik, Karl-Schwarzschild-Str. 1, 
D-85748 Garching bei M\"unchen, Germany\\
$^{2}$Apache Point Observatory, PO Box 59, Sunspot, NM 88349, USA}
\begin{document}

\date{Accepted . Received ; in original form 4 June 2003}

\pagerange{\pageref{firstpage}--\pageref{lastpage}} \pubyear{2003}

\maketitle

  \label{firstpage}

\begin{abstract}
We present a statistical analysis of halo emission for
a sample of 1047 edge-on disk galaxies imaged in five bands by the Sloan Digital Sky
Survey (SDSS). Stacking the homogeneously rescaled images of the galaxies,
we can measure surface brightnesses as deep as $\mu_r\sim31~\mathrm{mag~arcsec^{-2}}$.
The results strongly support the almost ubiquitous presence of stellar halos
around disk galaxies, whose spatial distribution is well described by a
power-law $\rho\propto r^{-3}$, in a moderately flattened spheroid ($c/a\sim 0.6$).
The colour estimates in $g-r$ and $r-i$, although uncertain, give a clear indication for extremely
red stellar populations, hinting at old ages and/or non-negligible metal enrichment.
These results support the idea of halos being assembled via early merging of satellite
galaxies. 
\end{abstract}


\section{Introduction}
The most commonly accepted paradigm for the formation of structure in the
universe, ranging from galaxies to clusters and superclusters of
galaxies, predicts that they are assembled via the hierarchical clustering of
dark matter (DM) halos, in the framework of the so called $\Lambda$CDM cosmology.
Although these models have been very successful in describing large scale
structure and much recent progress has been made in understanding 
the processes behind the assembly of individual galaxies through
numerical simulations (e.g. \citealt{navarro_white94}; \citealt*{sommerlarsen_etal99};
\citealt{navarro_steinmetz00}; \citealt{scannapieco_tissera03}),
semi-analytic modelling \citep*[e.g.][]{kauffmann_etal93,baugh_etal96,somerville_primack99},
and their combination (e.g. \citealt*{kauffmann_etal97}; \citealt{kauffmann_etal99,benson_etal00,springel_etal01}),
much work is still needed.\\
The study of the stellar halos of disk galaxies, such as our own Milky Way (MW), can give
a substantial contribution in this field. From many studies 
\citep[for a review see e.g.][]{majewski93}, it is known
that the MW halo is populated by old, metal-poor stars, whose origin is debated.
They may have formed in the early stages of the dissipational collapse of the gas in the
proto-galactic DM halo. Alternatively they may have been accreted through
the stripping of stars from satellite
galaxies. Evidence that this process plays an important role in building up 
the stellar population of the
halo, has come from the Sagittarius dSph \citep*{ibata_etal94}, 
a low-latitude halo stream found in the Sloan Digital Sky Survey \citep[hereafter SDSS,][]{yanny_etal03,ibata_etal03},
the ongoing disruption process of the Palomar 5 globular cluster \citep{odenkirchen_etal02},
and the detection of streams in local high velocity stars \citep{helmi_etal99}.
How much of the halo light and mass must be ascribed to this kind of interactions and
how common they were in the past is still unclear.\\
In order to answer these questions in a more general context, the extension of 
the observations to a statistical sample of external galaxies is required.
Unfortunately the surface brightness contributed by a stellar halo similar to
the one in the MW is typically 7 to 10 magnitudes fainter than the central
parts of the other galactic components (disk, bulge) and than the sky.
At present day, observations of the halo are available only for a handful 
nearby galaxies and provide varying results.
Recent work by \cite{ferguson_etal02} has demonstrated the
presence of substructure in the moderately metal-enriched stellar halo of M31, that
is likely to be the relic of the disruption of one or more small companion galaxy.
The search for halos in more distant galaxies, for which stars belonging to different
components cannot be resolved, has been focused on edge-on disks, because
of the much lower contamination of the halo by the projected 
disk stars. Problems with flat fielding, scattered light from other background and 
foreground sources, and extended PSF wings make it extremely difficult to
obtain reliable photometry down to 29-30 $\mathrm {mag~arcsec^{-2}}$, as required
in order to characterise galactic halos. After the first detection of halo
around NGC 5907 claimed by \cite{sackett94}, many controversial results have followed.
The latest and deepest observations of this galaxy \citep[see e.g.][]{zheng_etal99}
strongly support the extraplanar emission being instead a ring which results from tidal disruption 
of a satellite galaxy.
In the most complete study so far, (but see Sec. \ref{discussion_sec} for a 
more exhaustive review of recent results in the literature), the deep observations 
of a sample of 47 edge-on galaxies, by \cite{dalcanton_bernstein02},
provide evidence for the ubiquitous presence of red envelopes around disk galaxies, which
the authors attribute to an old, moderately metal-enriched, thick disk structure.\\

In this paper we present the results from the statistical study of halo
emission from a sample of more than 1000 edge-on disk galaxies imaged by the 
Sloan Digital Sky Survey \citep[SDSS, ][]{SDSS}. The SDSS is imaging about a
quarter of the sky in the $u$, $g$, $r$, $i$, and $z$ bands, with 54 sec drift scan exposures
at the dedicated 2.5 m Apache Point Observatory telescope
\citep{fukugita_etal96,gunn_etal98,hogg_etal01,smith_etal02,pier_etal03}, reaching 
$\sim25~\mathrm {mag~arcsec^{-2}}$ at $S/N\sim1$ for a single pixel.
In order to reach a surface brightness as low as $29~\mathrm {mag~arcsec^{-2}}$,
we adopt a stacking technique, in which we combine the images
of all the galaxies. First, the images must be geometrically transformed in order to
make the galaxies superposable. Other external sources must then be masked.
Finally, the count statistics of each pixel is considered and a suitable estimator
is chosen to represent the distribution. In this way it is possible
not only to increase the $S/N$ by a factor $\sim\sqrt{1000}$, but also
to remove statistically the major sources of contamination for
deep photometry of individual objects, namely foreground and background sources,
inhomogeneities in the flat field and light scattered inside the camera.\\
The paper is organised as follows: the sample is described in Sec. \ref{sample_sec};
in Sec. \ref{stacking_sec} we describe the image processing and the stacking
procedure. The results are analysed in Sec. \ref{analysis_sec}, paying
particular attention to the noise properties of the resulting images and to the
possible sources of error and of bias. A discussion, including a comprehensive 
comparison with previous results in the literature and the possible implications
for galaxy formation, is given in Sec. \ref{discussion_sec}.
A summary and the conclusions of this work are reported in Sec. \ref{conclusions}.


\section{The Sample}\label{sample_sec}
As of April 2002 the SDSS has covered $\sim 2000$ square degrees both in 5-band
imaging and spectroscopy. \cite{LSS} have selected a sample of
galaxies from the Main Galaxy Sample \citep{strauss_etal02}, also known as 
`Large-Scale Structure Sample 10' (LSS10), including all sources with average surface 
brightness within
the Petrosian\footnote{\cite{petrosian76} defines the Petrosian radius as the
radius at which the surface brightness equals a given fraction of the average surface
brightness inside that radius. The Petrosian flux (and hence the magnitude) is defined
as the flux inside a certain number of Petrosian radii. See \cite{EDR} for the full
description of the procedure adopted in the SDSS data reduction and the actual parameters.}
radius in $r$ band $\mu_r<17.77$, that have been successfully
targeted by the spectroscopic observations 
\citep[see][ for details about the `tiling' algorithm]{blanton_etal03}. 
We refer the reader to 
\cite{LSS} for the details of the sample and the regions of the sky
covered.\\
For the purposes of this work we selected a subsample of edge-on galaxies from
the LSS10 requiring the following conditions to be satisfied:\\
\begin{itemize}
\item{Petrosian magnitude ($Pmag$) successfully measured at least in the three most
sensitive SDSS pass-bands, namely the $g$, $r$, and $i$ band;}
\item{$Pmag_i \le 17.5$;}
\item{isophotal semi-major axis\footnote{the SDSS reduction pipeline PHOTO \citep{lupton_etal01}
provides an elliptical fit of the 25.0 $\mathrm {mag~arcsec^{-2}}$ isophote parametrised
by the semi-axes a and b} $a>10$ arcsec in $i$ band;}
\item{isophotal axis ratio $b/a \le 0.25$ in $g$, $r$ and $i$ band.}
\end{itemize}
Images of the 1221 selected galaxies were inspected by eye in
order to prune from the sample objects that are unsuitable for stacking.
First of all we discarded a few percent of the galaxies whose axis ratios had been
clearly underestimated due to some failure in the SDSS PHOTO reduction pipeline.
Galaxies showing evidence of interaction with nearby companions, warps 
or other irregularities were rejected. The absence of nearby bright sources
contaminating the background was required as well. The resulting sample
is composed of 1047 galaxies, ranging from $-22.5$ to $-16.0$ $i$-band absolute Petrosian
magnitude, in units of $\mathrm{mag}+5\log h$. 
The redshift distribution of the sample peaks at $z\sim 0.05$ with a standard 
deviation of 0.035. Typical physical dimensions range from $\sim1$ to $\sim25$ 
kpc $h^{-1}$ (Petrosian radius in the $i$ band), with a median value of 7.4 kpc $h^{-1}$.
\begin{figure}
\centerline{\includegraphics[width=9truecm]{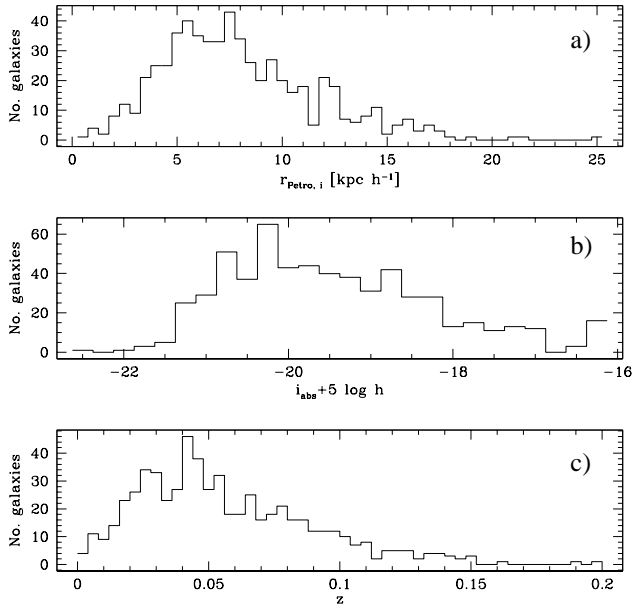}}
\caption{Distributions of the sample galaxies in
Petrosian radius (panel a), absolute $i$ band magnitude (panel b), and
redshift (panel c).}\label{statfig}
\end{figure}
A summary of the sample properties is reported in Fig. \ref{statfig}.
Eye inspection of the single galaxies confirms that our sample is dominated by late-type
disk galaxies, with at most a few per cents of Sb-c or Sb. Most of the galaxies appear to be
almost bulge-free and the the remainder have quite small `classical' bulges.

\section{The image stacking}\label{stacking_sec}
\subsection{The image pre-processing}
The raw SDSS imaging data is available as 2048x1489 pixel$^2$ (13.5x9.8 arcmin$^2$)
bias subtracted and flat-fielded `corrected frames'. Using SExtractor v2.2.2
\citep{sex} we accurately determined the centre and the position angle of
the major axis for each
galaxy. From the segmentation image output by SExtractor we obtain a mask for
all the sources in each field detected with S/N$\geq 1$.
The background level for each galaxy was determined as the mode of the pixel
count distribution in an annulus with inner radius $3\times a_{\mathrm{iso},i}$ 
and outer radius $6\times a_{\mathrm{iso},i}$ (where $a_{\mathrm{iso},i}$ is the isophotal major semi-axis
at 25.0 $\mathrm {mag~arcsec^{-2}}$ in $i$-band). This background was then subtracted
from each image.
\subsection{Photometric scale-lengths}
The main hypothesis underlying  the stacking technique as a reliable
statistical tool to investigate halo properties is an assumed 
self-similarity of the disk galaxies, or at least of their diffuse component.
There must exist a `characteristic' scale-length, such that
after rescaling, all galaxies can be superposed.
The validity of this assumption will be discussed in more detail
in Sec. \ref{profile_sect}.\\
Since the surface brightness distribution of edge-on disks is affected strongly 
by dust extinction, we chose the $i$ band ($\lambda_{\mathrm{eff}}=7480$\AA)
as the reference pass-band in order to characterise the scale-length of our
sample. This choice represents the best trade-off between the need for a high
sensitivity and the desire to limit the effects of dust\footnote{The $z$ band is in fact the red-most one, but its
sensitivity is about a factor 2-3 lower than the $i$ band.}.
Many different scale-lengths can be defined from photometric analysis.\\
We consider the following four scale-lengths:
\begin{itemize}
\item{the Petrosian radius $r_{\mathrm{Petro}}$, as obtained from the SDSS photometric
database \citep[see][]{EDR,DR1}}
\item{the effective or half light radius $r_{50}$}
\item{the exponential scale-length $r_{\mathrm{exp}}$}
\item{an isophotal radius obtained from the one-dimensional light profile
along the major axis $r_{\mathrm{1D}}$}
\end{itemize}
$r_{50}$, $r_{\mathrm{exp}}$ and $r_{\mathrm{1D}}$ have been determined using dedicated software
developed in the IRAF environment. In order to evaluate the two former
parameters, we derive
a surface brightness profile from circular aperture photometry extending out
to 1.5 isophotal ($\mu_i=25$ mag arcsec$^{-2}$) radii. $r_{50}$ is then obtained
as the radius enclosing 50 per cent of the total flux, while the exponential scale
radius $r_{\mathrm{exp}}$ is derived
from the least squares fit to the surface brightness profile between $r_{50}$
and $r_{90}$ (i.e. the radius enclosing 90 per cent
of the flux).\\
The basic motivation for analysing an one-dimensional brightness profile,
derived by collapsing the image of the galaxy along the minor axis, is to have
an estimate of the brightness which is as much as possible independent of the
inclination. The surface brightness enhancement is, in fact, important and severely
dependent on the inclination, when the edge-on condition is approached. Assuming
low or negligible extinction, the quantity $\zeta\equiv -2.5 \log \frac{\mathrm dFlux(x)}{x\mathrm dx}$,
where $Flux(x)$ is the flux enclosed within the projected distance $x$ from the
minor axis of the galaxy, is independent both of the inclination and of the 
distance of the galaxy. As reference level for determining $r_{\mathrm{1D}}$ we chose 
$\zeta=25.0~\mathrm{mag~arcsec^{-2}}$, which corresponds roughly to 
a surface brightness of $\mu=25.0~\mathrm{mag~arcsec^{-2}}$
for a typical edge-on galaxy.\\
All four scale-lengths are quite strongly correlated, as expected for self-similar
exponential profiles. The typical scatter around the 1:1 relation between 
pairs of corresponding scale-lengths is $\sim 15$ per cent.

\subsection{The image transformation and stacking}
\begin{figure*}
\centerline{\includegraphics[width=9truecm]{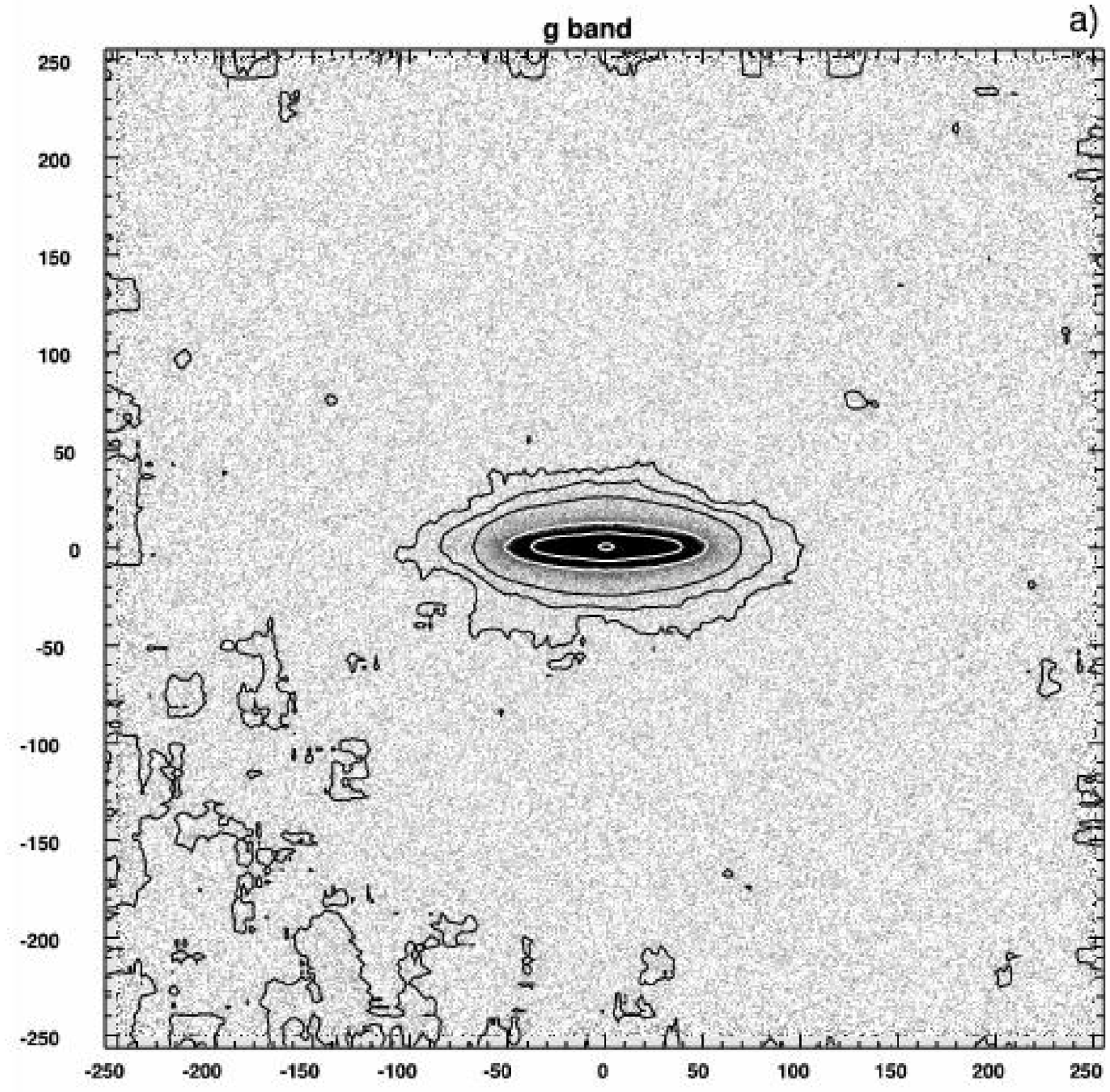}
\includegraphics[width=9truecm]{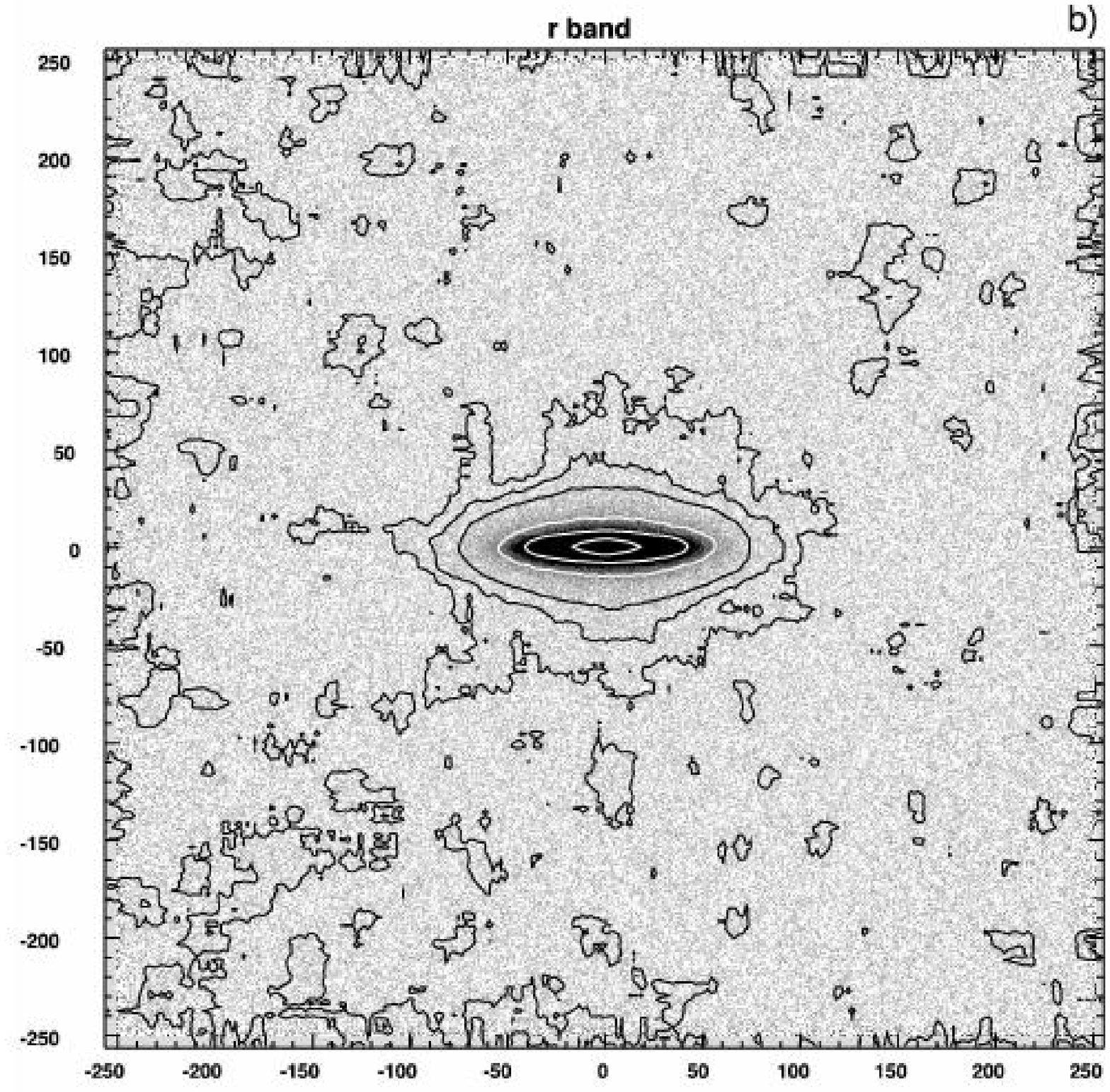}}
\centerline{\includegraphics[width=9truecm]{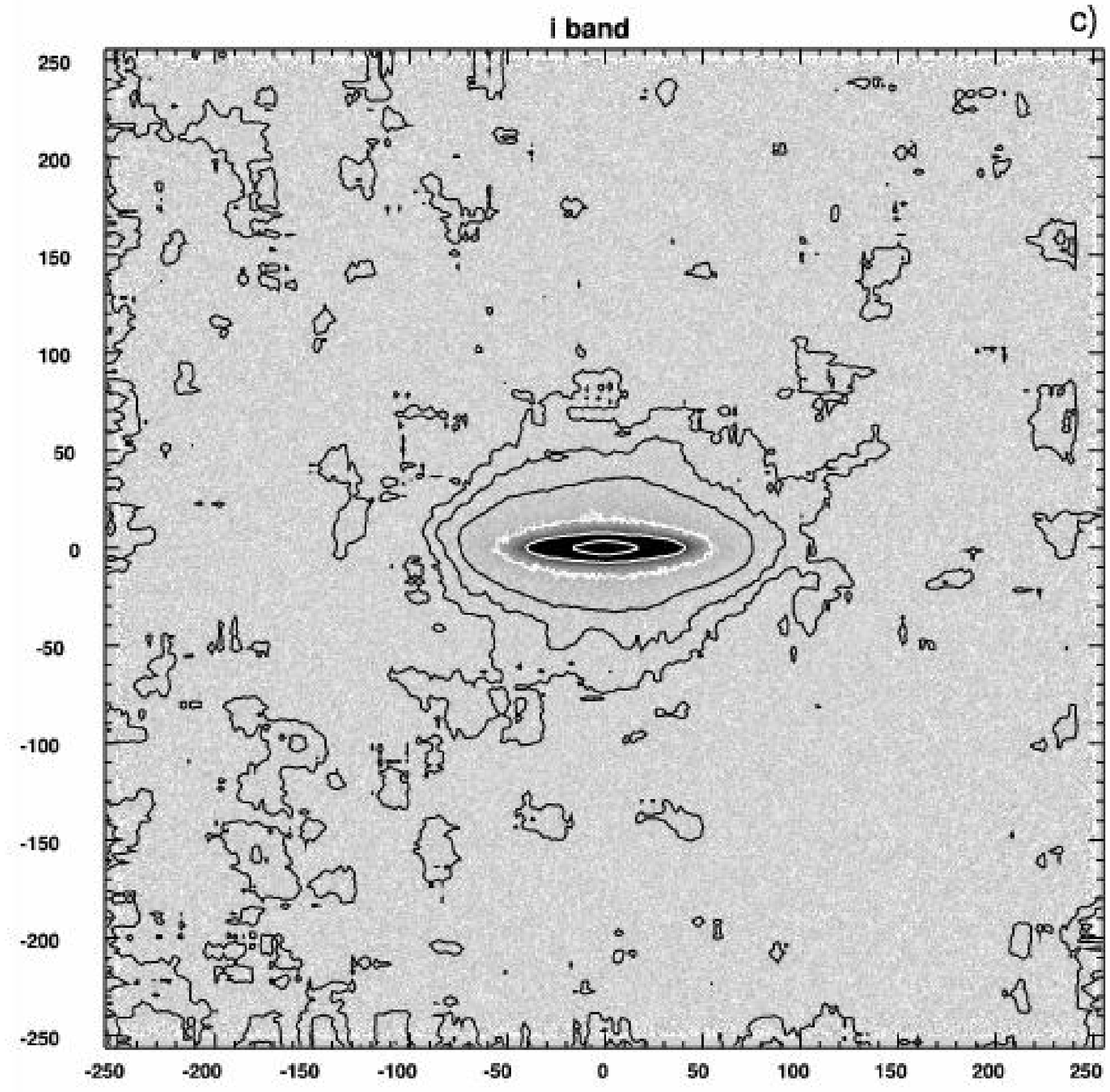}
\includegraphics[width=9truecm]{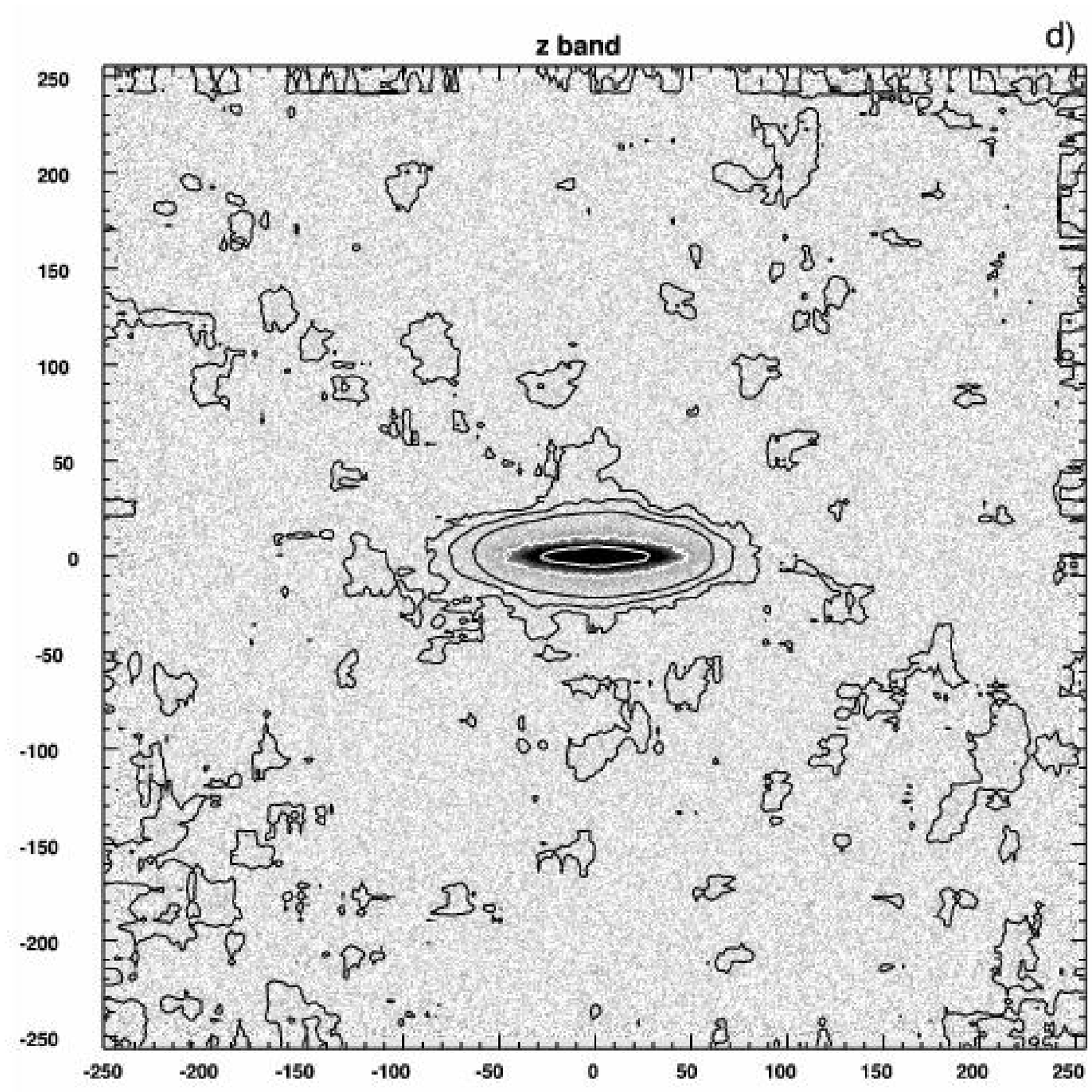}}
\caption{The images resulting from the stacking of the galaxies rescaled
according to their exponential scale length, in $g$, $r$, $i$, and $z$ band. 
Intensity is coded in grey
levels through a square root transformation, adopting arbitrary bias and contrast.
The overplotted isophotal levels are as
follows: 30.0, 29.0, 28.0, 26.0, 24.0, 22.0 for the $g$ and
the $r$ band; 29.5, 28.5, 27.5, 25.5, 23.5, 21.5 for the $i$
band; 28.0, 27.0, 26.0, 24.0, 22.0 for the $z$ band. See text
for the details. The scales indicate the offset from the galaxy centre in pixel,
10 pixel=$r_{\mathrm{exp}}$.}\label{contourfig}
\end{figure*}

For each of the four scale-lengths considered, the median value in the sample, rounded to 
an integer number of pixels\footnote{The pixel scale of the SDSS CCDs is $0.396~\mathrm{arcsec~pix^{-1}}$},
has been taken as reference for the rescaling, as given in Table \ref{resc_table}.
\begin{table}
\caption{Median scale-lengths (in arcseconds) and rescaling reference values 
(in pixel)}\label{resc_table}
\begin{tabular}{cc | cc | cc | cc}
\hline
\multicolumn{2}{c|}{$r_{\mathrm{Petro}}$}&\multicolumn{2}{c|}{$r_{50}$}&
\multicolumn{2}{c|}{$r_{\mathrm{exp}}$}&\multicolumn{2}{c|}{$r_{\mathrm{1D}}$}\\
arcsec &pix&
arcsec &pix&arcsec &pix&arcsec &pix\\
\hline
10.07&25&4.74&12&3.76&10&15.50&40\\
\hline
\end{tabular}
\end{table}
All galaxy images have been geometrically transformed using the drizzle re-sampling
technique (as implemented in the IRAF {\bf geotran} task) in order to correctly
propagate the original noise properties. Each image
has been translated to the galaxy centre, rotated according to the measured position
angle of the major axis, and expanded or contracted according to the ratio between the
reference scale length and the corresponding measured one. The same transformation has
been applied to the corresponding mask, and a `grow radius' of  10 pixels (in the
original image size) around each masked pixel has been adopted in order to mask
the extended PSF wings of the brightest sources.
Since each image has a different photometric calibration and the measured surface
brightness of a galaxy is differently affected by Galactic attenuation, depending 
on its location in the sky, an intensity rescaling is needed as well. The transformation from
the original pixel intensity $i$ to the rescaled one $i'$ is given by the
following formula:
\begin{displaymath}i'=i\times 10^{-0.4\left(Z_{\mathrm{p}}-Z_{\mathrm{p,ref}}-K \times Airmass-Reddening\right)},
\end{displaymath}
where $Z_{\mathrm{p}}$ and $Z_{\mathrm{p,ref}}$ are the flux calibration zero points of the image and of
reference, respectively, $K \times Airmass$ is the airmass correction, and $Reddening$ is
the attenuation due to our Galaxy, as given by \cite*{schlegel_dust}. All the quantities
are expressed in magnitudes.\\
\\
The combination of the transformed and rescaled images has been performed using
the {\bf imcombine} IRAF procedure. First of all, the masked pixels are rejected. The
median of the count distribution of the remaining pixels is then calculated in order
to obtain the median image. An average image is calculated as well, after
clipping the 16 per cent percentile tails of the count distribution of the unmasked
pixels. Adopting the standard approximation for the mode of a distribution
\begin{displaymath}mode=3\times median - 2 \times average,
\end{displaymath} we calculate the mode image.
The resulting $512\times512~\mathrm{pixel}^2$ images extend much beyond
the detected emission from the galaxies and allow us 
to determine the properties of the background in great detail. 
A careful comparison between the three
statistical combinations shows that, despite of the conservative masking
method adopted, the diffuse luminosity of sources other than the considered
galaxies can result in significant skewness in the pixel count distribution
and in a systematic increase of the sky surface brightness up to 
$31~\mathrm{mag~arcsec^{-2}}$ in i-band in the outermost regions of the
average frame with respect to the innermost ones, in which contaminating
sources are avoided by selection. This systematic effect in the background
is almost completely removed in the mode image. In the following, therefore, we
will always refer to that one as the resulting stacked image.\\
For each stacked image the residual background level is determined as the count mode
in an annulus of 130 pixel inner radius, 96 pixel thick, and subtracted.\\
In Fig. \ref{contourfig} we present the images for the four most sensitive
pass-bands ($g$, $r$, $i$, $z$), obtained by stacking the images rescaled according
to the galaxy exponential scale length. For the $u$ band the 
signal-to-noise is
not sufficient to say anything about the presence and the characteristics of
halo emission, therefore we will neglect this band in the following analysis and discussion.
Rescaling according to the other scale-lengths does not change the resulting
images significantly, as we will demonstrate more quantitatively later on,
and therefore they are not shown here.
Intensity levels in Fig. \ref{contourfig} are coded in square-root transformation
grey scale, suitably adjusted to show the maximum extension of the low surface 
brightness envelope. We superpose isophotal contours in two magnitude ranges:
black contours, corresponding to the faintest isophotes, are obtained adopting
a boxcar smoothing scale of $10\times10$ pixel, whereas the white, brightest ones, have been
calculated with a $2\times2$ pixel smoothing scale. The represented isophotal levels are 
given in the figure caption.
The images and the superposed isophotes show very clearly the presence of a diffuse
luminous envelope around the disk, in all four considered bands. This halo
(whatever physical meaning we give to this word) is particularly evident
and round in the $r$ and the $i$ bands, whereas it is significantly flattened in
the $g$ and $z$ bands (but the lower sensitivity in the latter band does not allow us to
say much).

\section{The analysis}\label{analysis_sec}
In this section we present the results of an analysis of the stacked images
in the $g$, $r$, $i$, and $z$ pass-bands. We mainly concentrate on the stacking of the
complete sample (1047 galaxies), because of the resulting higher sensitivity,
but we also consider the stacking of three subsample of bright, intermediate
and low luminosity flat galaxies (Sec. \ref{lbin_sect}),
in order to better understand the possible
 dependence of the observed halo properties on total luminosity.
After analysing the background properties of the stacked images in order to
assess our detection limits (Sec. \ref{noise_sect}), the photometric 
properties are investigated by means of radial sector-averaged
surface brightness profiles (Sec. \ref{profile_sect}), and compared to different
models of luminosity distributions, namely thin+thick disks and disk+halo (Sec. 
\ref{models_sect}). The average halo colours are presented in Sec. \ref{colours_sect}.
Unless otherwise specified, we will always use the stacking of the images
rescaled according to the exponential disk scale-length. We devote an entire paragraph
in Sec. \ref{profile_sect} to demonstrate that the adopted scale-length is not critical
in determining the characteristics of the observed halos.

\subsection{The background noise properties}\label{noise_sect}
Since our photometric measurements are performed by integrating the flux 
over different image areas, as will be explained in Sec. \ref{profile_sect},
it is crucial to determine the noise properties on different
scales. The presence of large-scale fluctuations is already obvious
from a superficial inspection of the contour plots in Fig. \ref{contourfig}.
We restrict the analysis of the noise to the annulus (130 pixel inner radius
and 96 pixel thickness) on which the sky level has been computed, and we consider
the rms of the intensity after re-binning the image by different linear factors, ranging
from 1 to 50. The dependence of the $rms$ on the scale-length $L$ is well described
by a power law of the form 
\begin{equation}
rms(L)=rms_0~L^{-\alpha}\label{bkgnoise},
\end{equation}
where $rms_0$ is the noise on one-pixel scale, and $\alpha\sim0.7-0.8$, thus
indicating that the noise has significant large scale structure in excess to what
is expected in the case of pure Gaussian noise (for which $\alpha=1$).

\subsection{The surface brightness profiles}\label{profile_sect}
\begin{figure*}
\centerline{\includegraphics[width=8truecm]{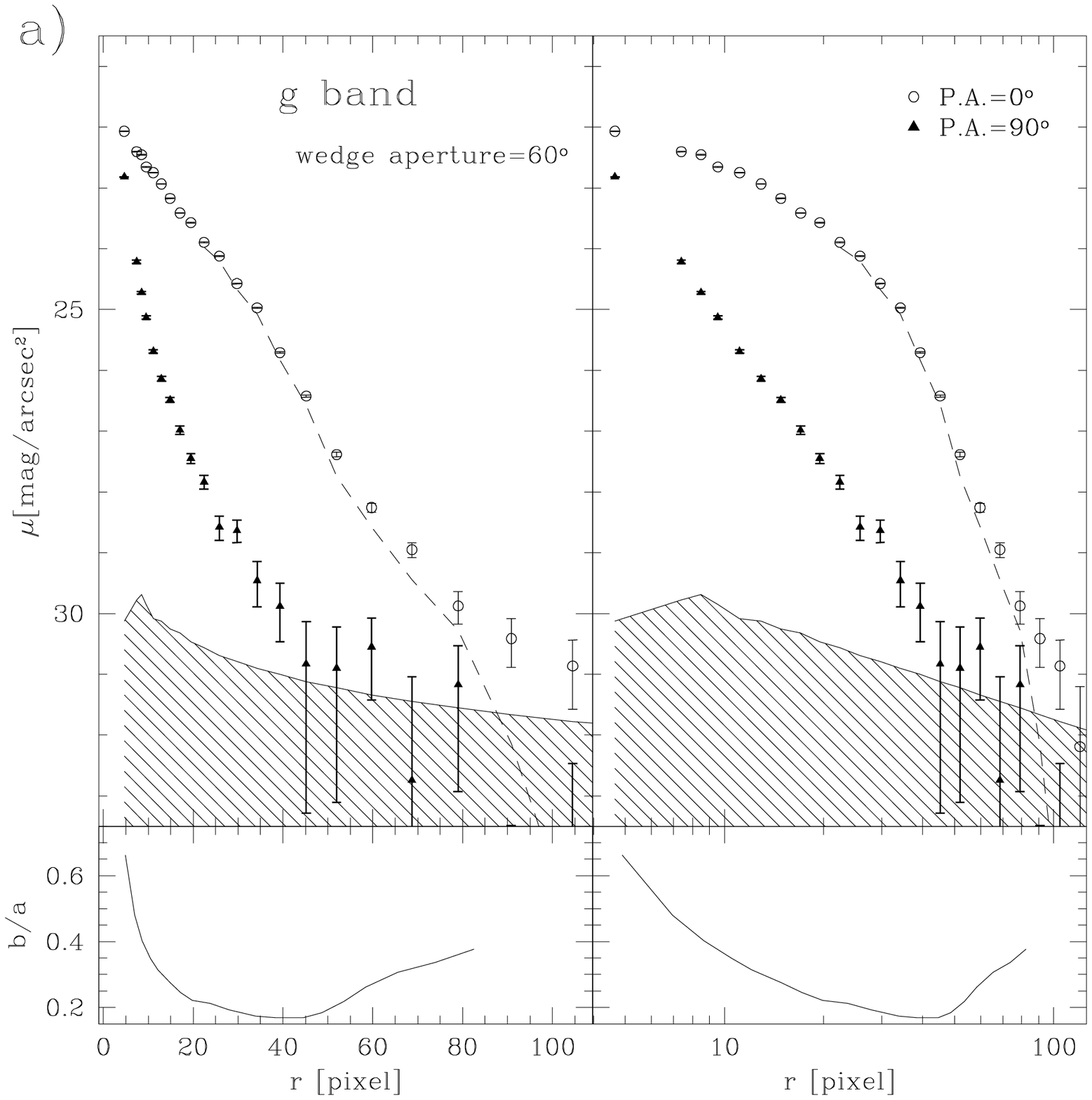}
\includegraphics[width=8truecm]{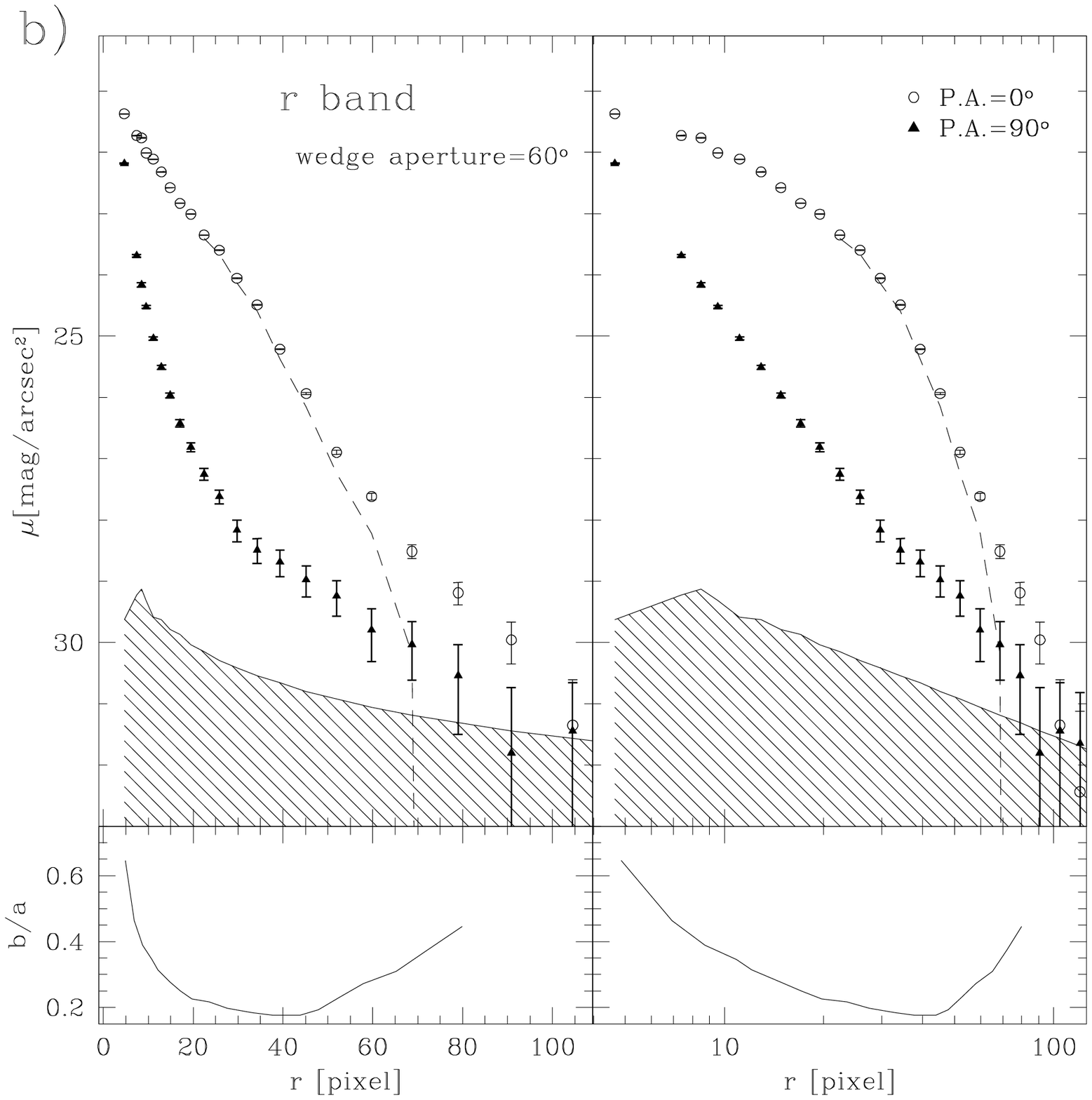}}
\centerline{\includegraphics[width=8truecm]{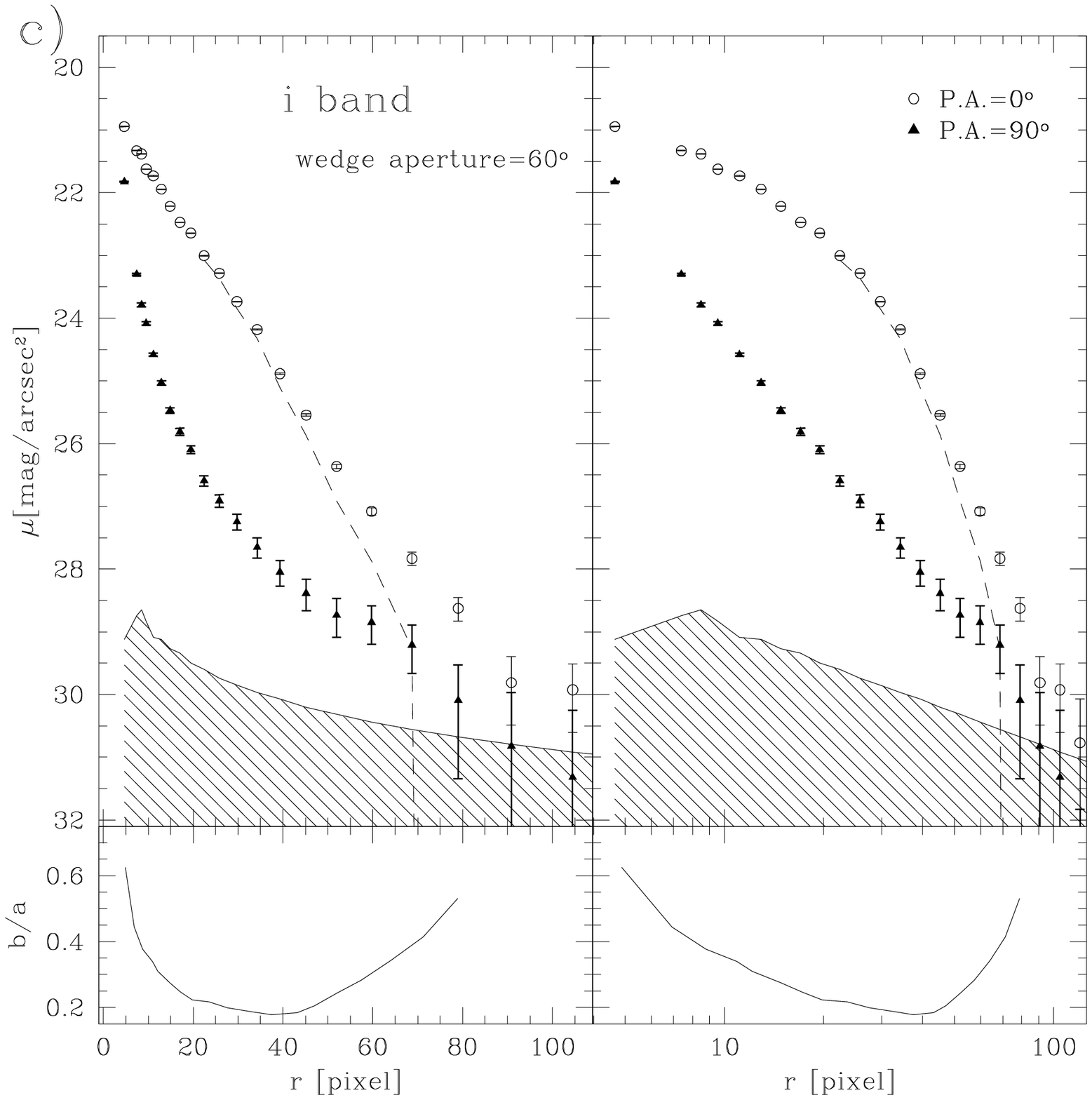}
\includegraphics[width=8truecm]{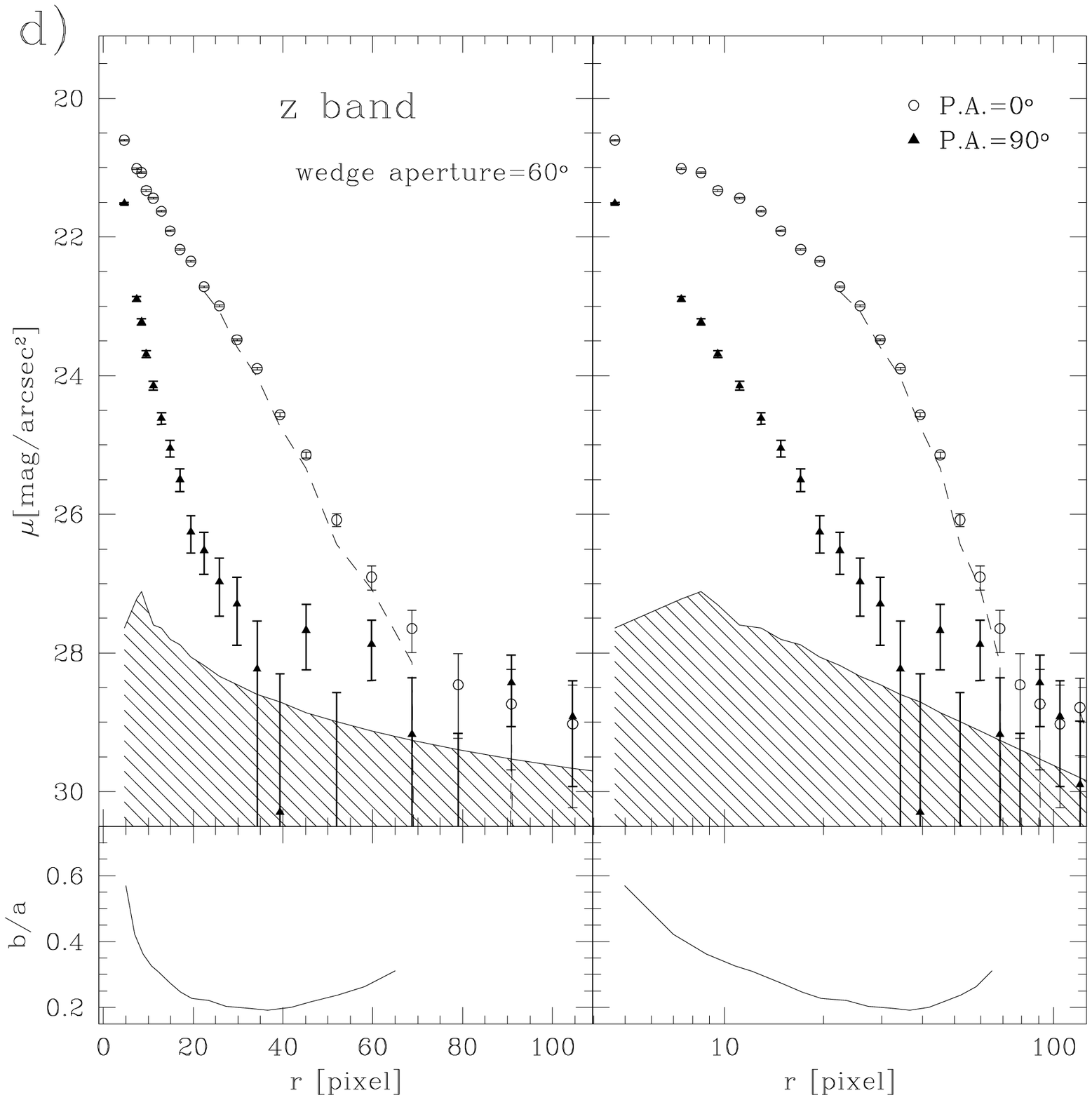}}
\caption{The SB profiles in $g$, $r$, $i$, and $z$ band, in $60^{\mathrm o}$ wedges along the disk 
(PA=$0^{\mathrm o}$, open circles) and perpendicular to it (PA=$90^{\mathrm o}$, filled 
triangles). Distances from the centre in pixels (10 pixels corresponding to $1~r_{\mathrm{exp}}$)
are shown in linear scale (left panels) and logarithmic scale (right panels). 
The sky rms is represented by the shaded areas. Dashed lines represent the
`pure' disk profile at $r>20$ pix, obtained as explained in the text. The bottom 
panels of each graph show the $b/a$ axial ratio of the isophotes as a function of the
major axis $a$.}\label{sbprof_fig}
\end{figure*}

The most obvious way to investigate the properties of the approximately round,
low surface brightness structures emerging in Fig. \ref{contourfig} is to extract
surface brightness (SB) profiles, averaging the flux in large wedges at different position
angles. First of all, we divide the image in four circular sectors of 
$60^{\mathrm o}$ aperture, 
centred at $0^{\mathrm o}$, $90^{\mathrm o}$, $180^{\mathrm o}$ and $270^{\mathrm o}$
position angles (PAs). Each of these sectors is in turn radially divided into a
number of coronae, geometrically spaced such that the outer radius of the
$k$th corona is given by $r_k=r_0\cdot 1.15^k$, with $r_0=6~\mathrm{pixel}= 0.6~r_{\mathrm{exp}}$.
Finally, we estimate the mean SB as a function of the radius for  
$0^{\mathrm o}$ (i.e. along the disk) and $90^{\mathrm o}$ PA (i.e.
perpendicular to the disk) averaging the SB in pairs of
corresponding coronae in the two symmetric sectors at $0^{\mathrm o}$ and $180^{\mathrm o}$
and at $90^{\mathrm o}$ and $270^{\mathrm o}$ PA, respectively.
In each of the four graphs of Fig. \ref{sbprof_fig} we show the SB profiles
for the $g$, $r$, $i$, and $z$ images respectively, in linear
scale (upper left panel) and logarithmic scale (upper right panel): open circles represent
the $0^{\mathrm o}$ PA profile, filled triangles the $90^{\mathrm o}$ PA.
The shaded areas represent the level of the rms background fluctuations at the
scale length corresponding to the area of the coronae in which the SB is averaged, as
calculated in the previous section. Thus the $1-\sigma$ detection limits for the SB
profiles can be assessed as $\sim 31~\mathrm{mag~arcsec}^{-2}$ for the $g$, $r$ and $i$ band
and as $\sim 28~\mathrm{mag~arcsec}^{-2}$ for the $z$ band.\\

In order to evaluate the errors on the SB we must consider contributions from
1) the background fluctuations and 2) the intrinsic scatter in the average
signal from a galaxy. The first one is just provided by the background rms as 
evaluated
in the previous section using Eq. \ref{bkgnoise}, with $L$ given by the square
root of the area of the corona over which the signal is integrated.
The estimation of the intrinsic scatter contribution is much more cumbersome, since
the shape of the statistical distribution of the pixel counts in the coronae
is not known \emph{a priori} and the light distribution of the galaxy itself is likely
to give a major contribution to the rms of the count statistics.
In order to remove it, for each pixel we consider the deviation from the average intensity
of the corresponding pixels in the four image quadrants, which are symmetric with respect to the
 x and y axes, and express the intrinsic rms as:
\begin{equation}
rms_{\mathrm{intrinsic}}=\sqrt{\frac{1}{3~N}\sum_{i=1}^{N}{\sum_{j=1}^4\left[ I_{ij}-\left< 
I_{ij} \right>_j\right]^2}}\label{rms_int}
\end{equation}
where $I_{ij}$ is the pixel intensity, $N$ is the number of pixels per corona per quadrant,
and $\left< I_{ij} \right>_j$ is the average intensity over the $j$ index.
In the absence of any large scale background fluctuation the error on the average intensity is
just obtained dividing the (\ref{rms_int}) by $\sqrt{4 N}$. Adding the contribution of the
large scale fluctuations, we can write the error as:
\begin{equation}
\sigma=\sqrt{\frac{\left(rms(L)\right)^2}{2}+\frac{rms^2_{\mathrm{intrinsic}}}{4 N}}
\end{equation}
where $rms(L)$ is given by Eq.\ref{bkgnoise} with $L$ given by the square
root of the area of the corona, and the factor 2 comes in from having averaged two coronae.
Such an error estimate has been proved to be consistent with the scatter of the average
intensity in the corresponding coronae in the four quadrants.\\

In the lower panels of each graph in Fig. \ref{sbprof_fig} we plot the axial
ratio $b/a$ of the isophotes as a function of the radius (semi-major axis).
This is calculated as follows.
We assign a number of SB levels and determine the radius $r$ at which such SB's
are reached at different PAs $\theta$, by interpolating the SB profiles extracted from 
6 wedges per quadrant. Assuming $r=r(\theta)$ is well represented by an ellipse with
the major axis along the x direction, we derive the best fitting semi-axes $a$ and $b$
by means of a standard least squares algorithm.\\

\begin{figure}
\centerline{\includegraphics[width=9truecm]{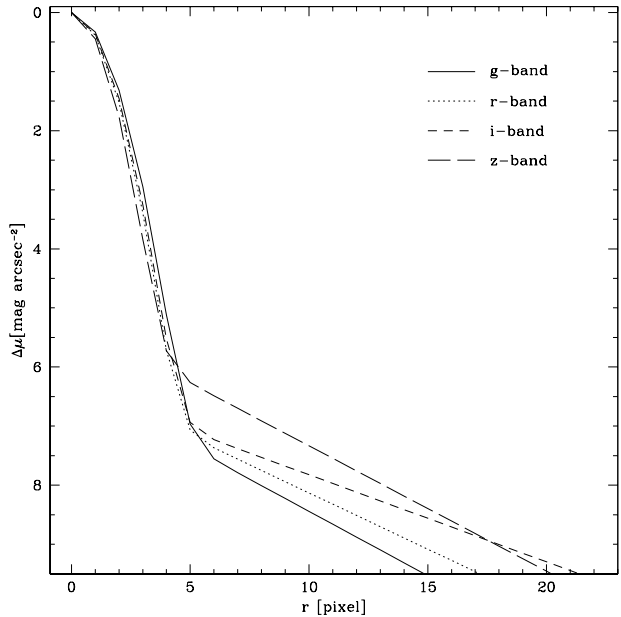}}
\caption{The PSF profiles in $g$ (solid line), $r$ (dotted line), $i$ (short-dashed line), and $z$ band
(long-dashed line), expressed as difference of magnitude with respect to the central surface
brightness. The lines represent the analytic fit (Gaussian core plus exponential wings) to the 
radial profile measured on the stacked image of stars, as explained in the text.}\label{PSF}
\end{figure}

As it is apparent from the straight-line behaviour of the open circles in the linear-scale 
plots of Fig. \ref{sbprof_fig}, the profile of the wedges centred on the disk show an 
exponential decrease of the SB with radius, as it is typical for disks. Perpendicular to the 
disk, instead, the rapid initial decrease of the SB becomes shallower as we go further away 
from the centre of the galaxy. The trend is well approximated by a power-law with index 
$\sim 2$, as the straight-line behaviour of the triangles in the log-scale plots points out.
This is in general true for the four bands analysed. However, the relative intensity at the
same radial distance along the disk and perpendicular to it is different in different
bands, reflecting a possible dependence of the flattening of the diffuse halo on the band
and, in turn, this points to colours gradients in the halo itself.
As already noted from the isophotal contours in Fig. \ref{contourfig}, the halo is prominent
in the $r$ and the $i$ band, reaching a SB comparable to that of the disk at 
$r\sim 8- 10~r_\mathrm{{exp}}$ and making the isophotal shape significantly rounder
($b/a\sim 0.5- 0.6$). In the $g$ band the halo has very little extension (the surface brightness
drops below $31~\mathrm{mag~arcsec^{-2}}$ beyond $\sim 4~r_{\mathrm{exp}}$), thus leaving the isophotes
extremely flattened. For the $z$ band the measurements are inconclusive: there is
some hint of a shallowing of the slope of the profile, but the $S/N$ drops below 1
already at $\sim 3~r_{\mathrm{exp}}$ at the level of $28.5~\mathrm{mag~arcsec^{-2}}$.\\
\\
In Fig. \ref{sbprof_fig}, for $r>20$ pix, we also show as dashed lines the `pure disk'
profiles
obtained subtracting the $90^{\mathrm o}$ profile from the $0^{\mathrm o}$ profile,
after stretching the former by a factor $0.6^{-1}$ to take the flattening of the halo
into account (see Sect. \ref{models_sect} for details on the 0.6 factor).
There is evidence that the `pure' disk slope steepens beyond $4-5~r_{\mathrm{exp}}$
in the $r$ and $i$ bands, consistent with studies of individual disks in the literature
\citep*[see e.g.][ and references therein]{vanderkruit01,kregel_etal02}, whilst
the results are unclear in the $g$ and $z$ bands.\\
\\
In order to rule out scattered light and the extended wings of the point spread 
function (PSF) as major contributors
to the observed halos, we have performed the same stacking procedure on the images of stars
taken from the same frames as the galaxies to generate effective PSF's in each band.
In each frame we identify a star 1) whose central brightness differs by less than
$2~\mathrm{mag~arcsec^{-2}}$ from the central SB of the galaxy and 2) which is distant from
other contaminating sources, as requested
for the galaxy selection. Then, sources other than the star are masked, the frame is re-centred
on the star and the same geometrical transformations as applied to the galaxy are performed.
The radial profile of the PSF can be reproduced by a Gaussian core plus exponential wings over
a large extension. 
Fig. \ref{PSF} shows the analytic fit to the measured PSF's in the four bands, expressed as the
difference of magnitude with respect to the central surface brightness. The decline in the core
is very sharp, $6- 7~\mathrm{mag~arcsec^{-2}}$ within 5 pixel, and the exponential wings
contribute less than $1/10^4$ of the central surface brightness at radii larger than 20 pixels.
We will analyse the effects of the PSF on the measured profiles in the next section, by convolving
with different models for the light distributions.
\\

\begin{figure*}
\centerline{\includegraphics[width=8truecm]{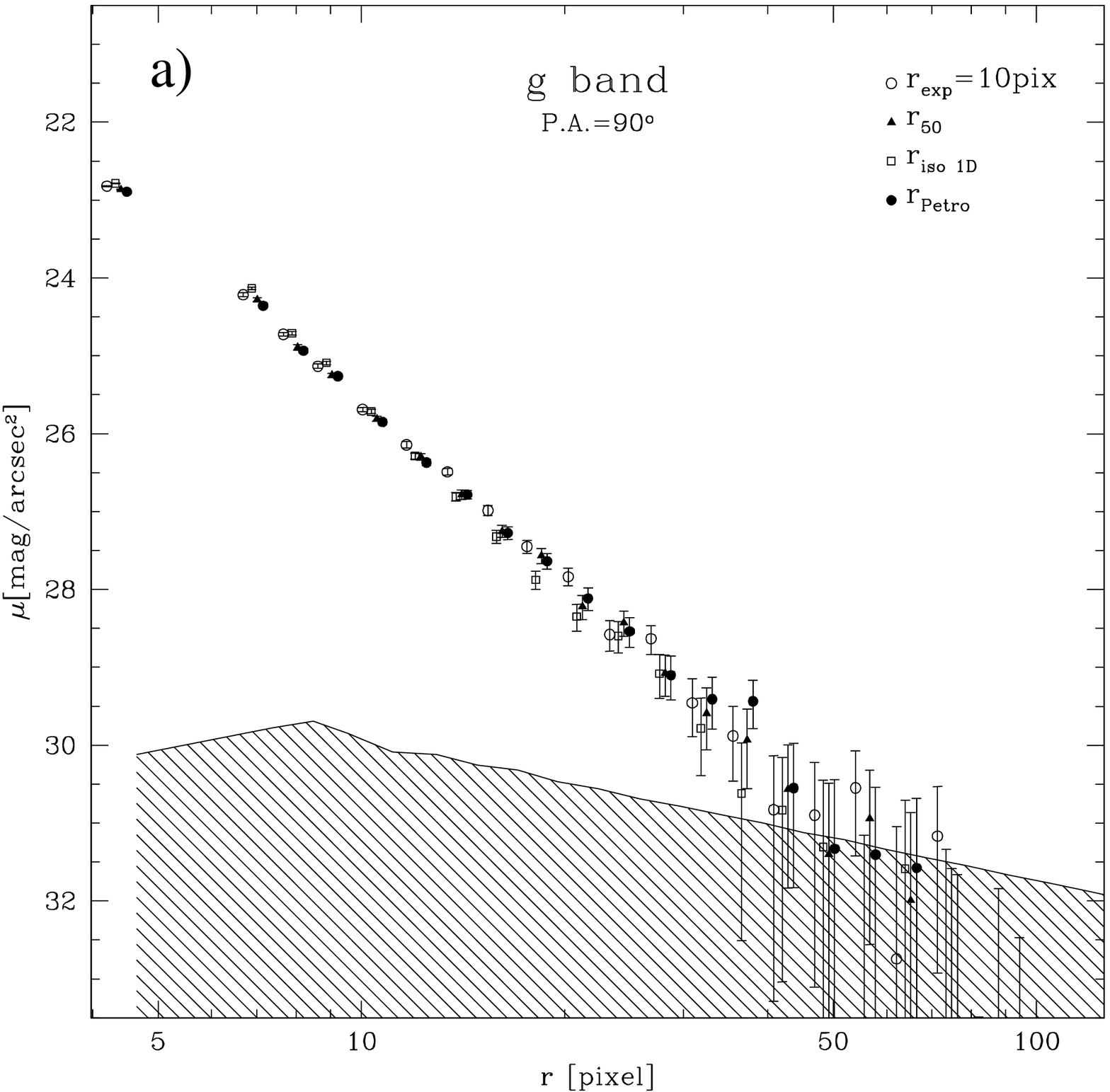}
\includegraphics[width=8truecm]{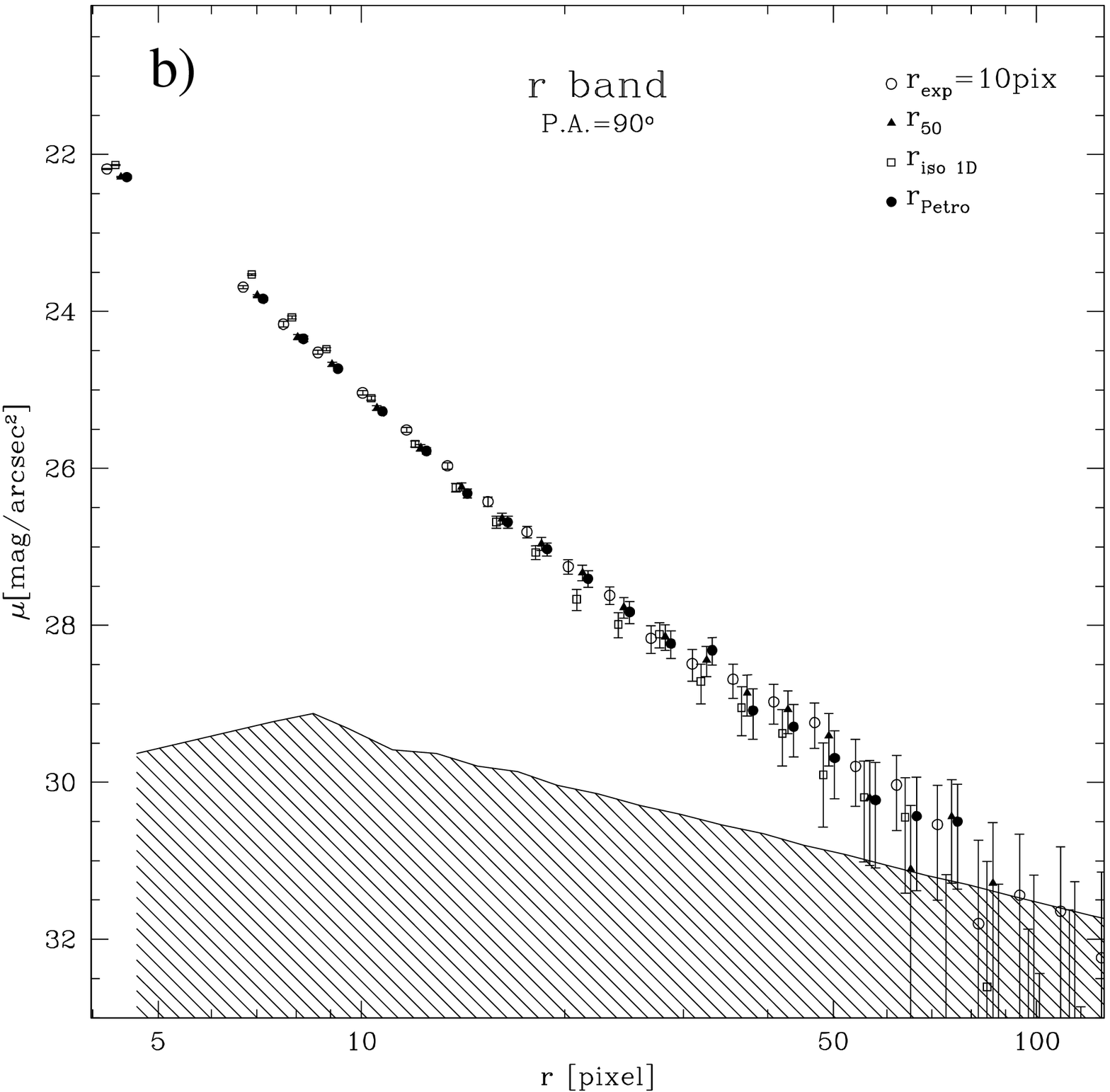}}
\centerline{\includegraphics[width=8truecm]{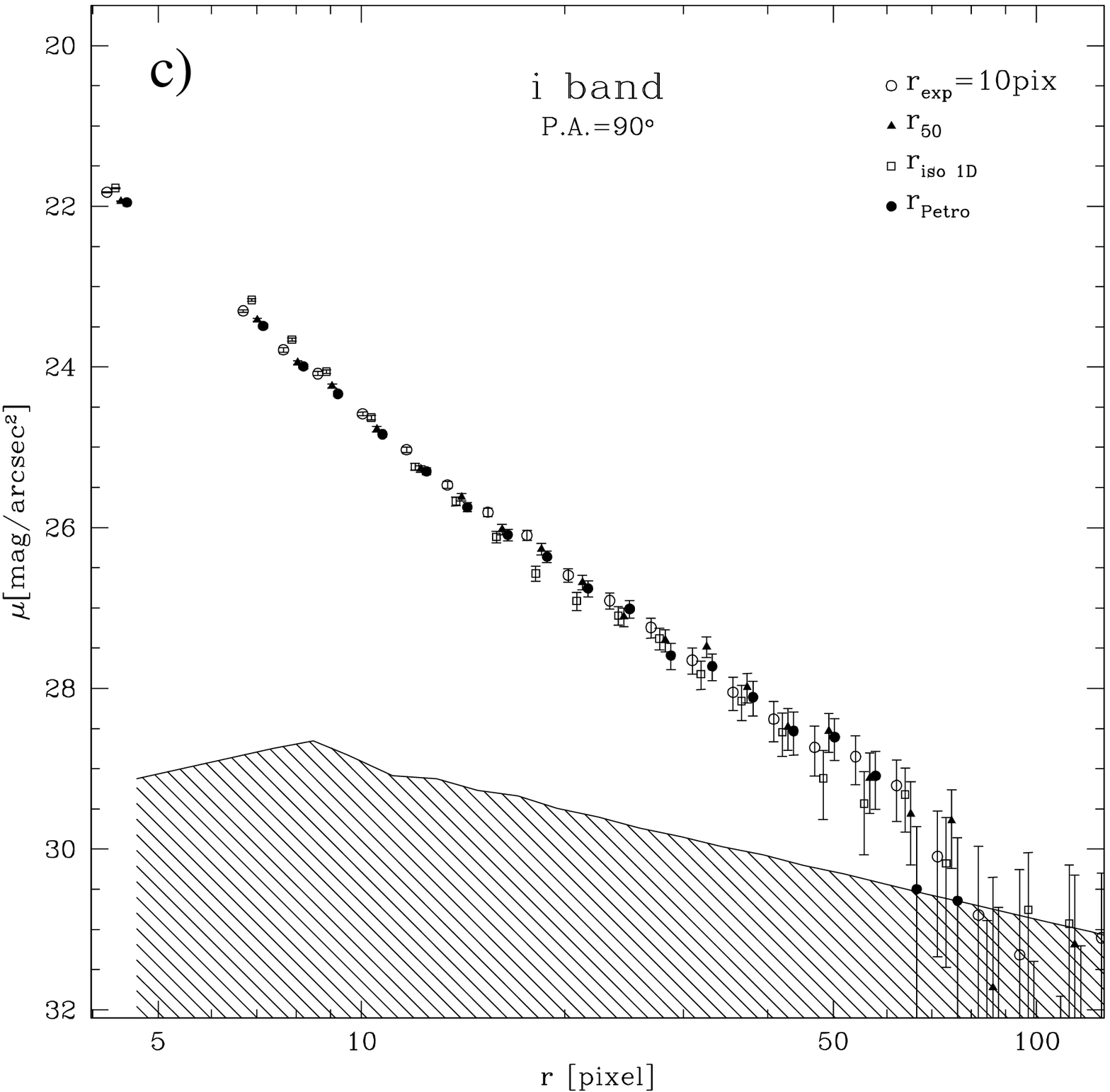}
\includegraphics[width=8truecm]{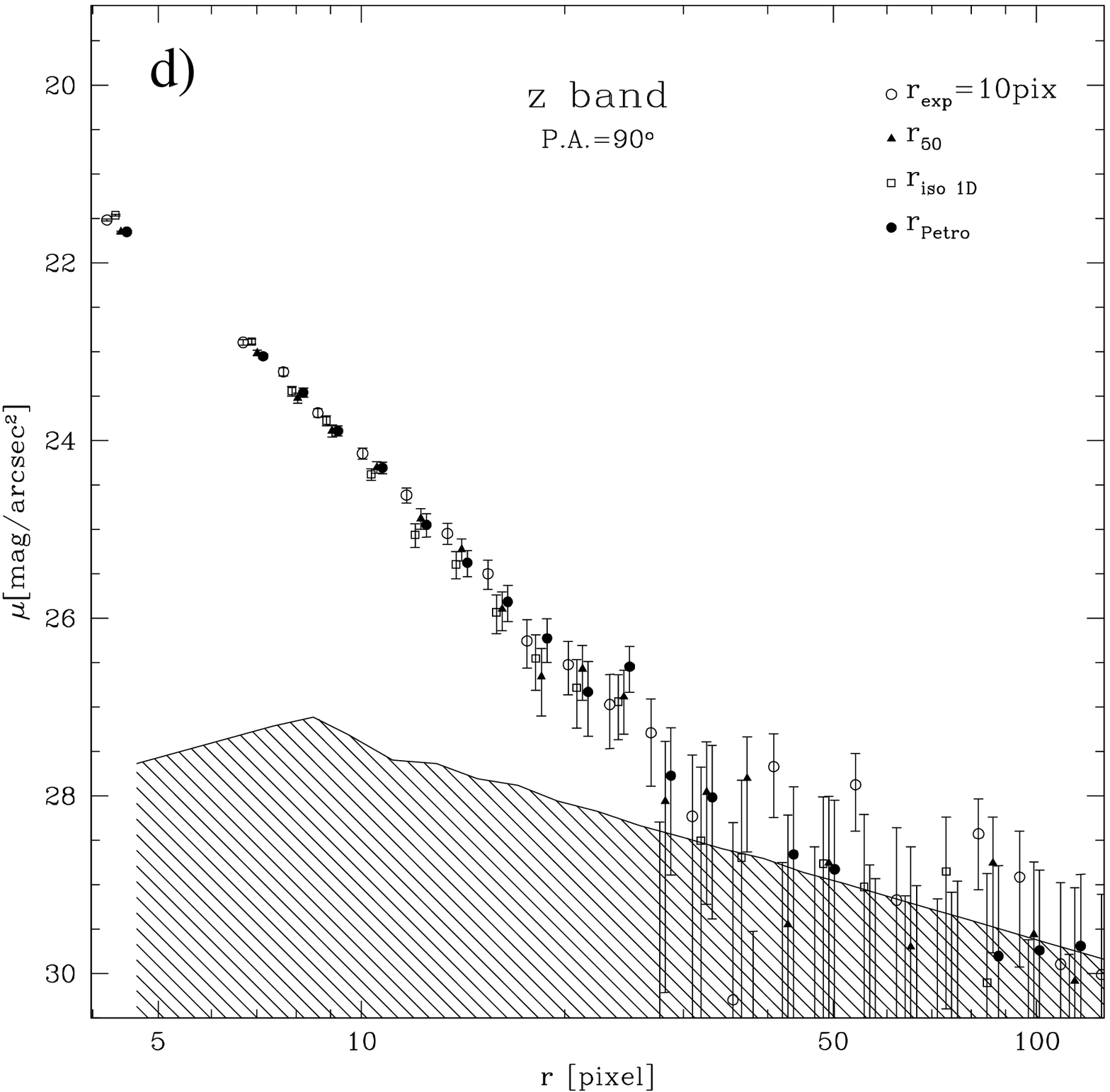}}
\caption{The surface brightness profiles in $g$, $r$, $i$ and $z$ band, at $90^{\mathrm o}$ PA,
for different adopted rescalings. Open circles refers to images rescaled according to
$r_{\mathrm{exp}}$, filled triangles to $r_{\mathrm{50}}$, open squares to $r_{\mathrm{iso 1D}}$ and
filled circles to $r_{\mathrm{Petro}}$. The sky rms is represented by the shaded areas.}\label{cfrescale}
\end{figure*}

The characteristics of the observed halo profiles show little or no dependence on the scale-length
adopted for rescaling the images of the galaxies. In Fig. \ref{cfrescale} we plot the SB profile
in the $90^{\mathrm o}$ PA wedge in the four bands, using different symbols for the
different rescalings:
open circles for $r_{\mathrm{exp}}$, filled triangles for $r_{50}$, open squares for $r_{\mathrm{iso 1D}}$ and
filled circles for $r_{\mathrm{Petro}}$. The $r$ coordinate of each point is exactly rescaled in order to 
match the sample median values of the four scale-lengths considered. As in Fig. \ref{sbprof_fig},
the shaded areas represent the sky rms. The error-bars are calculated as explained before.
The agreement between the different rescalings is extremely good: almost all points are consistent
within the error-bars, with a handful of exceptions for the $r_{\mathrm{iso 1D}}$ rescaling. Even for these,
the deviation is less than 3 $\sigma$. Thus we conclude that there is no systematic dependence
of the average halo properties on the profile shape of the disk as described by variations
in relative scale-lengths.

\subsection{Modelling the halo}\label{models_sect}
In this section we investigate the structural properties of the detected emission by means of
simple models of the light distribution. In particular, we consider the possible contribution
from a thick disk component, with an exponential vertical light density distribution, and from
a (moderately) flattened power-law halo. Making predictions from models of this kind 
is non-trivial and must take into account that the observed emission results from a 
double convolution of the `true' average light density distribution of the galaxies with 1) the
distribution of inclinations and 2) the effective PSF, as computed above.
The models are calculated as follows. First we assume a particular 3-dimensional
distribution of light. We produce a set of 1000 Monte Carlo realizations of the projected surface
brightness, uniformly varying the inclination angle of the disk between 0 and 15 degrees (roughly
corresponding to a projected axial ratio between 0 and 0.25 for an infinitely thin disk,
as required by the sample selection criterion for the galaxies). The 1000 realizations are then
averaged and convolved with the analytic PSF computed in the previous section, separately for each
band.
We do not expect to reproduce the stacked images near the nucleus nor at small distances from
the disk plane, because we do not model dust extinction. Thus we compare our 
models to the observations by means of vertical-cut profiles, that allow us to exclude these 
`forbidden' regions. In each cut, whose width is chosen to be proportional to the distance 
from the minor axis of the image, we average the flux coming from the four quadrants.\\

In the thin+thick disk model the 3-dimensional light density of each individual galaxy is
assumed to be given by:
\begin{equation}
\nu(r,z)=N e^{-\frac{r}{r_{\mathrm{exp}}}} \left[e^{-\frac{|z|}{z_{\mathrm{thin}}}}+R\frac{z_{\mathrm{thin}}}{z_{\mathrm{thick}}}
e^{-\frac{|z|}{z_{\mathrm{thick}}}} \right]
\end{equation}
where $N$ is the normalisation factor, $R$ is the flux ratio of the thick over the thin disk,
$r_{\mathrm{exp}}$ is the exponential scale-length of the radial SB distribution, $z_{\mathrm{thin}}$ and $z_{\mathrm{thick}}$ are the
exponential vertical scale-lengths of the thin and thick disks respectively. We fix
$r_{\mathrm{exp}}\equiv 10$ pixels, and $z_{\mathrm{thin}}\equiv 1$ pixel. No disk truncation is adopted.
The realizations cover
a logarithmic spaced grid of four by four values in the $z_{\mathrm{thick}}-R$ parameter space 
($z_{\mathrm{thick}}$ ranging from $10^{0.2}$ to $10^{0.8}$ pixels, $R$ from 0.01 to 0.4). The
normalisation $N$ is left as a free parameter and fitted by minimising the $\chi^2$.
We find that the thin+thick disk models fail to reproduce the observed SB distribution of the
observed halos, since they predict a sharp exponential cut-off
of the SB along the $z$ coordinate, which is inconsistent with the observed power-law profile.\\
\begin{figure*}
\centerline{\includegraphics{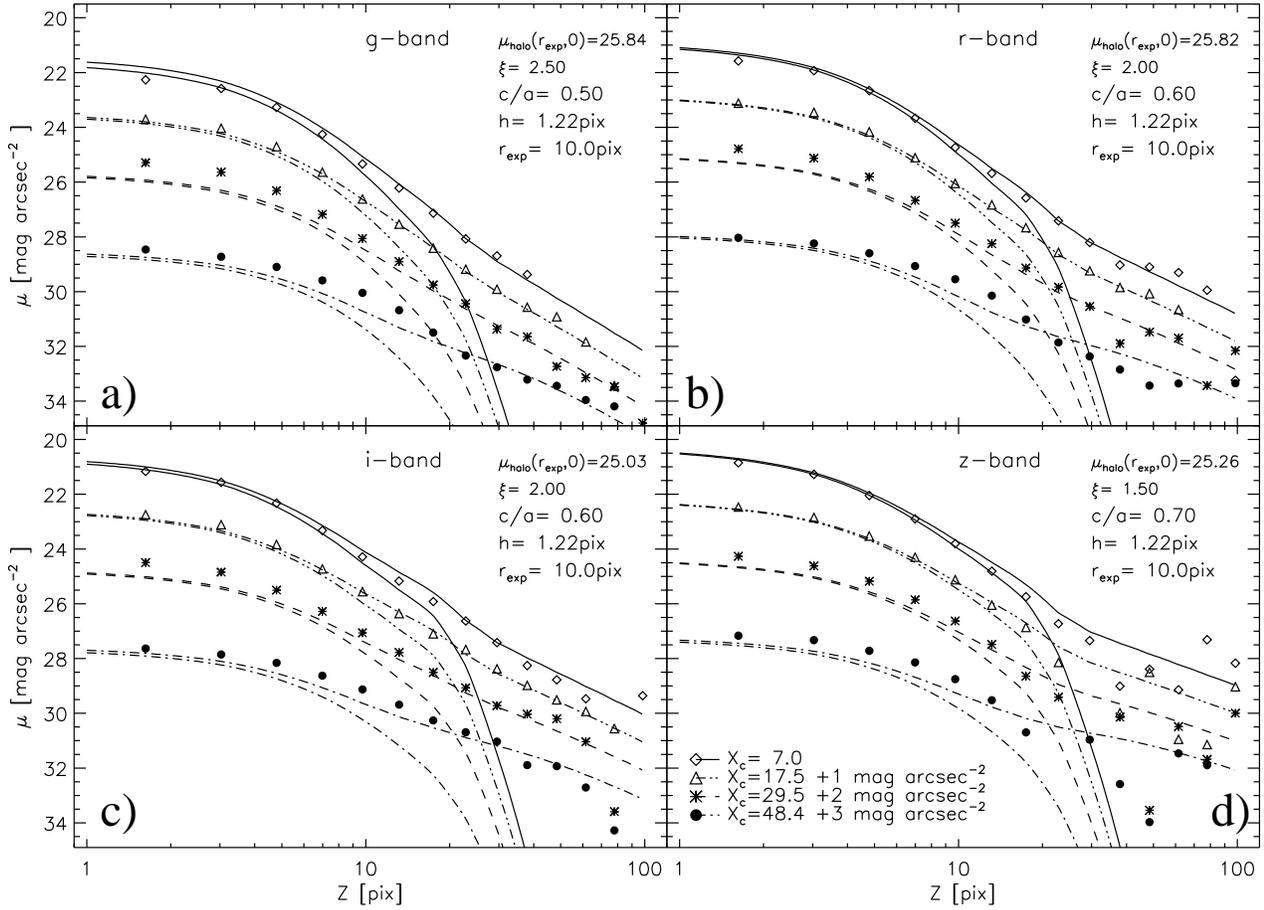}}
\caption{Vertical cut SB profiles with superposed the prediction from the power-law halo+disk model.
Points are the SB measured in four vertical cuts at distance $X_c$ from the centre (in pixel units, 
see legend). Profiles are offset $1~\mathrm{mag~arcsec^{-2}}$ from each other. Heavy lines 
represent the model profiles, light lines are the disk component of the model 
alone.}\label{dhmodels}
\end{figure*}
In the disk+halo models, we assume a halo component with a generalised Hubble density distribution
as introduced by \cite*{halo_powerlaw}, modified to allow the iso-density surface to be oblate 
spheroids:
\begin{equation}
\nu_{\mathrm{halo}}(r,z)\propto\left[1+\frac{r^2+\frac{z^2}{(c/a)^2}}{r_c^2}\right]^{-(\xi+1)/2}
\end{equation}
where $c/a$ is the flattening parameter, $r_c$ is the softening parameter or core radius\footnote{
The core radius is introduced only for mathematical convenience, to avoid the central divergence
of a pure power law.}, and $\xi$ is the power-law index of the projected SB.
The disk component is modelled by a double exponential distribution in $r$ and $z$, without any
truncation:
\begin{equation}
\nu_{\mathrm{disk}}(r,z)\propto e^{-\frac{r}{r_{\mathrm{exp}}}} \cdot e^{-\frac{|z|}{h}}
\end{equation}
with $r_{\mathrm{exp}}$ and $z_0$ representing the exponential scale lengths in $r$ and $z$ respectively.
We fix $r_c\equiv 1$ pixel and $r_{\mathrm{exp}}\equiv 10$ pixels, and realize a grid of models in the
$h-\xi-(c/a)$ parameter space, covering the following ranges: $h=0.50- 3.00$,
$c/a=0.3- 1.0$, $\xi=1.50- 4.00$. The total (disk+halo) normalisation and the normalisation
of the halo relative to the disk  for each model on the grid have been fitted by minimising 
the $\chi^2$.
The best fitting models for the $g$, $r$, $i$ and $z$ band are represented in Fig. \ref{dhmodels}. For
each band we plot the vertical SB profiles at four different distances $x_c$ from the centre
of the galaxy, as obtained in the vertical cuts described above, offset by
$1~\mathrm{mag~arcsec^{-2}}$ one from the other. Even if the $\chi^2$ are extremely high, thus
demonstrating that the adopted models cannot reproduce in the details the complexity of the 
galaxy structure, the general agreement with the measured points is satisfactory and we are
successful in reproducing the trend of the profiles.
Besides the total model profiles (heavy lines), we plot also the exponential disk components alone
(light lines): while dominating at small height, they give negligible contribution at $z \gtrsim 30$
pixels ($3~r_{\mathrm{exp}}$).
The best fitting model parameters are reported in the panels of Fig. \ref{dhmodels}. The disk
scale-height is quite well constrained to $h \sim \frac{1}{10}r_{\mathrm{exp}}$. The power-law slope ($\xi$)
is steeper in $g$ band (2.50), and increasingly shallower in $r$ and $i$ (2.00) and $z$ band (1.50), while
the halo shapes get increasingly rounder from $c/a=0.50$ in $g$, to 0.60 in $r$ and $i$, and 0.70 in 
z band. This is in good agreement with the previous analysis on the images themselves
and on the radial profiles.
The surface brightness of the halo component at $r=r_{\mathrm{exp}}$ along the minor axis is
25.84, 25.82, 25.03 and 25.26 $\mathrm{mag~arcsec^{-2}}$ in the $g$, $r$, $i$ and $z$ band respectively.
We can estimate the amount of halo light coming from outside the 
$25~\mathrm{mag~arcsec^{-2}}$ isophote as $\sim 2-3$ per cent
of the total galaxy light.\\
The models derived above allow us to quantify the pollution of the outer envelope by scattered 
light and PSF wings. We note that, at $r=20$ pix, the PSF-convolved disk component
contributes $\sim30$ per cent of the total surface brightness\footnote{20 per cent in the $g$, 33 per cent in the $r$ and
the $i$, and 40 per cent in the $z$ band}. This contribution decreases
very rapidly at larger distances, and becomes negligible at $r>30$ pix in all the bands.
We conclude that scattered light from the disk
contributes much less than 30 per cent of the total (disk+halo) measured light beyond 20 pixel.
\subsection{The halo colours}\label{colours_sect}
\begin{figure}
\centerline{\includegraphics[width=9truecm]{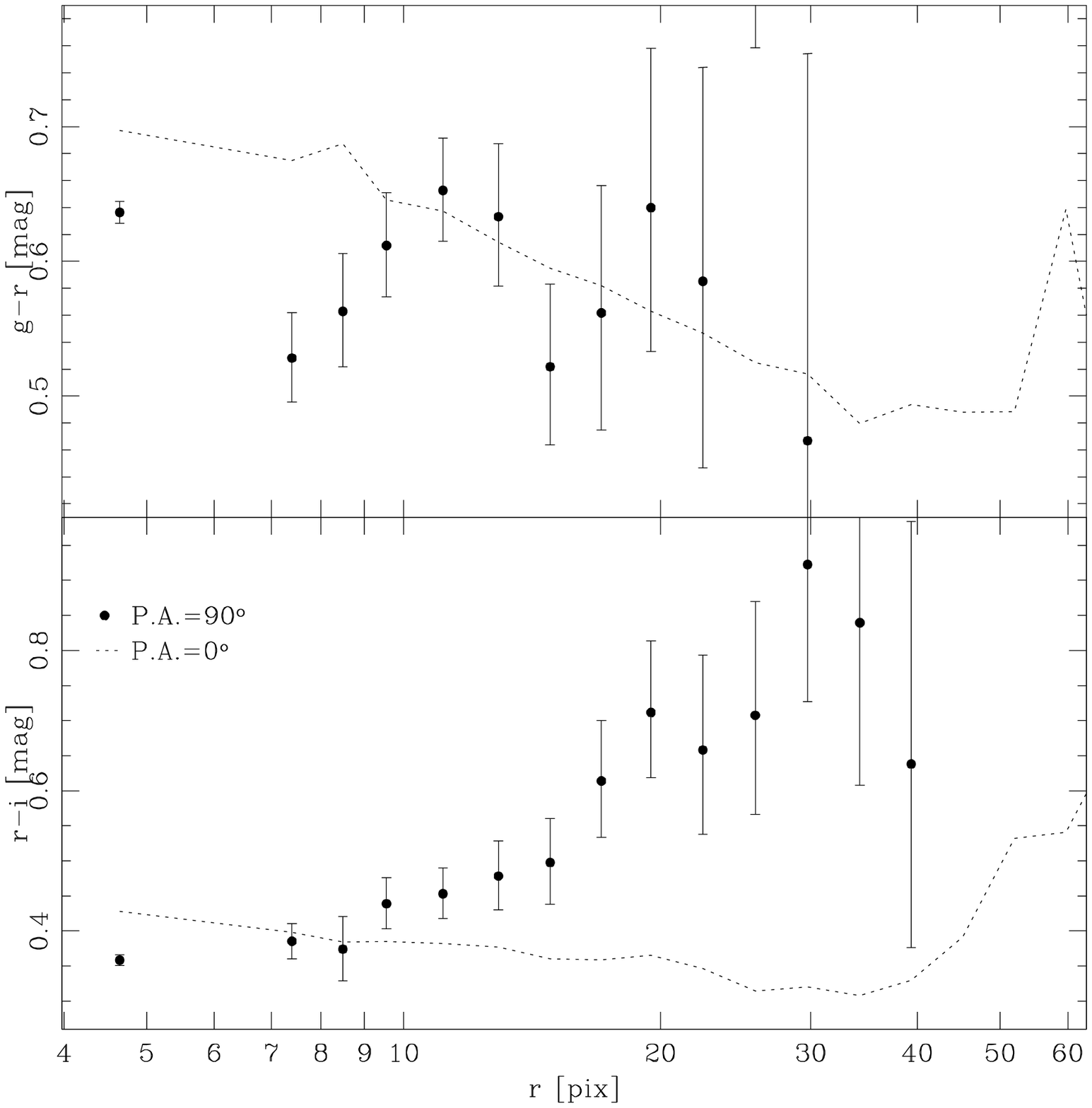}}
\caption{The $g-r$ and $r-i$ colour profiles. The filled circles refer to the $60^{\mathrm o}$-aperture
sector perpendicular to the disk, the dotted lines represent the colours in the wedge along
the disk.}\label{colour_plot}
\end{figure}

Based on the SB profiles presented in Sec. \ref{profile_sect}, we derive the colour profiles in
the two $60^{\mathrm o}$-aperture sectors including the disk and perpendicular to it. We concentrate
on the $g-r$ and $r-i$ colours, excluding colours involving the $z$ band, because of its lower sensitivity.
In Fig. \ref{colour_plot} the dots represent the colours measured perpendicular to the 
disk, with the error-bars derived from the errors on the SB profiles; for comparison, 
with the dotted lines we plot the
colour profiles for the disk. This has a blueing gradient toward the outer parts, which
is particularly evident for $g-r$, but it is still clearly apparent in $r-i$. Disk colours 
($g-r=0.7- 0.5$, $r-i=\sim 0.4$) are consistent with typical star-forming galaxies,
once we allow for significant dust extinction, as expected for edge-on conditions.
The increasing presence of dust near the galaxy centre and age and metallicity gradients 
in the disk can also easily explain the observed gradients.
Beyond $4-5~r_{\mathrm{exp}}$ there is evidence for colours getting redder, possibly indicating
that the halo is overtaking the disk. This is consistent with the steepening of the
profiles of the `pure disk', shown as dashed lines in Fig. \ref{sbprof_fig}.\\
The $g-r$ profile perpendicular to the disk is extremely noisy: no clear trend can be established
and the measurements are definitely unreliable beyond 20 pixels. The halo in these bands is roughly
as red as the inner parts of the disk.
The $r-i$ colour shows instead a clear reddening toward the outskirts of the galaxy, reaching
extremely red colours $\sim 0.8$, about 0.4 mag redder than the disk.
We derive the best estimate of the halo colours using the mean colour around 20 pixels
and consider the uncertainties given both by the error bars and by the scatter of the
points around an `ideal', smooth profile:\\
$g-r=0.65\pm0.1$\\
$r-i=0.60\pm0.1$\\
These results are not corrected for the effects of the PSF. Based on the analysis in
the previous section, we estimate that the SB at $r=20$ pixel is polluted by scattered light
from the inner parts by up to $\sim30$ per cent. However, this contribution is very similar in all the
bands (20 per cent in the $g$, 33 per cent in the $r$ and the $i$ bands) and can redden the real halo colours 
by some hundredths up to 0.2 mag (worse case for $g-r$). It is worth stressing
that this assessment has large uncertainties, due to the very simplistic nature of
the models here adopted and to their extreme sensitivity to the fitting parameters (e.g.,
changing the vertical scale-length of the disk affects very slightly the goodness of the
fit, but can significantly change the disk and scattered light contributions to
the colours at 20 pixels).\\
The robustness and the significance of the estimates given above, 
will be discussed in detail in Sec.\ref{discussion_sec}.

\subsection{Dependence on the total galaxy luminosity}\label{lbin_sect}
\begin{figure*}
\centerline{\includegraphics[width=8truecm]{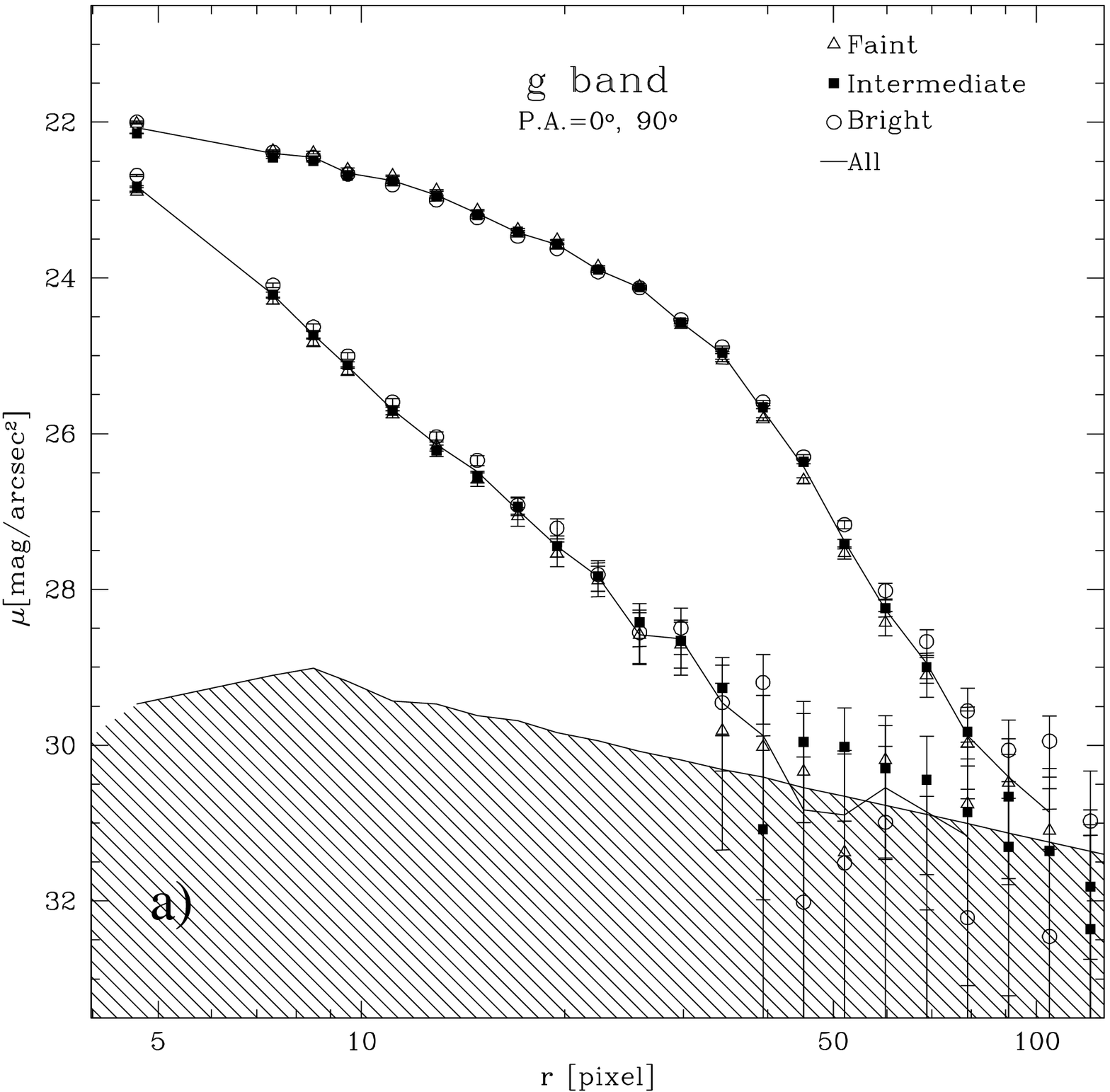}
\includegraphics[width=8truecm]{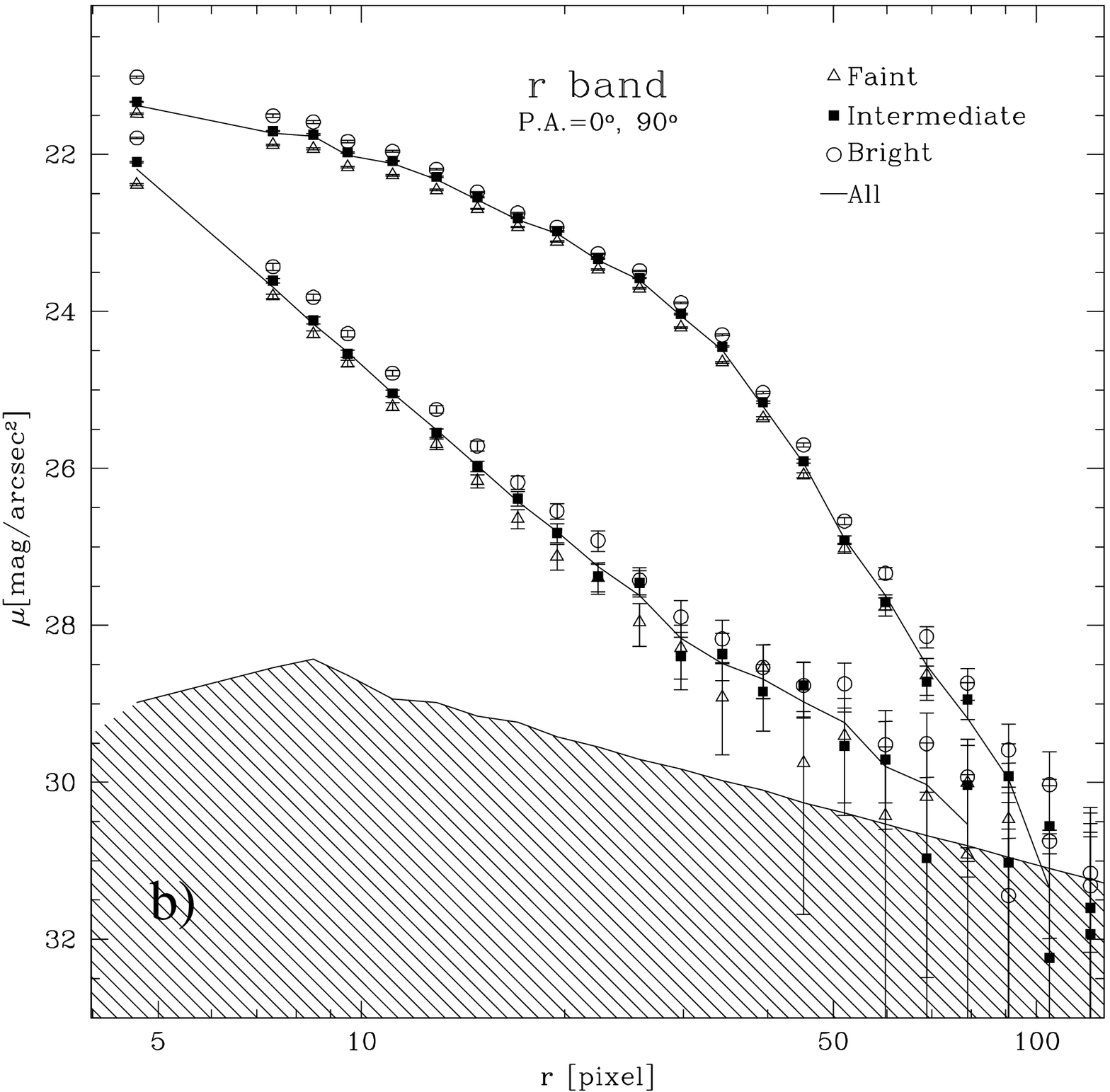}}
\centerline{\includegraphics[width=8truecm]{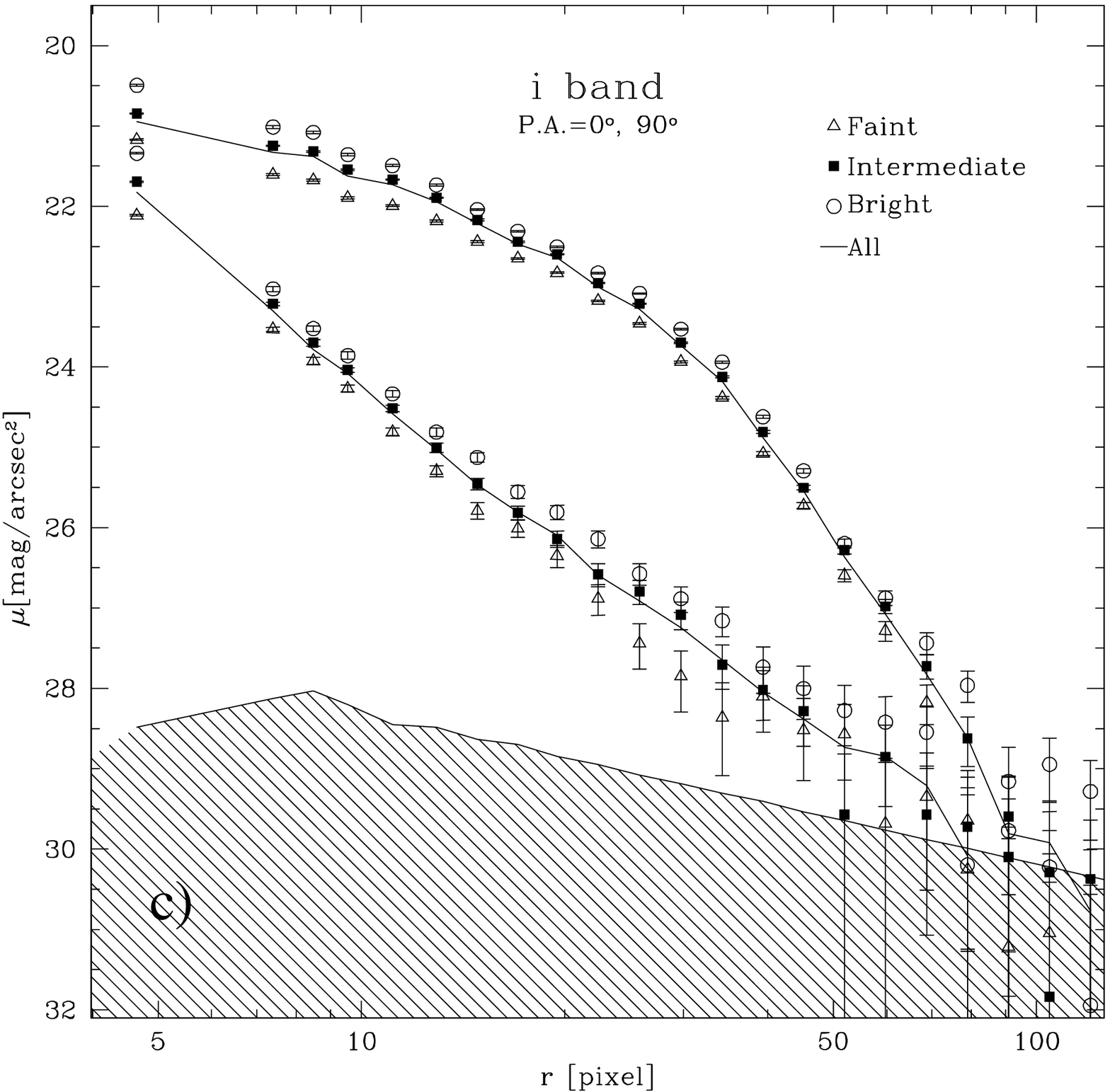}
\includegraphics[width=8truecm]{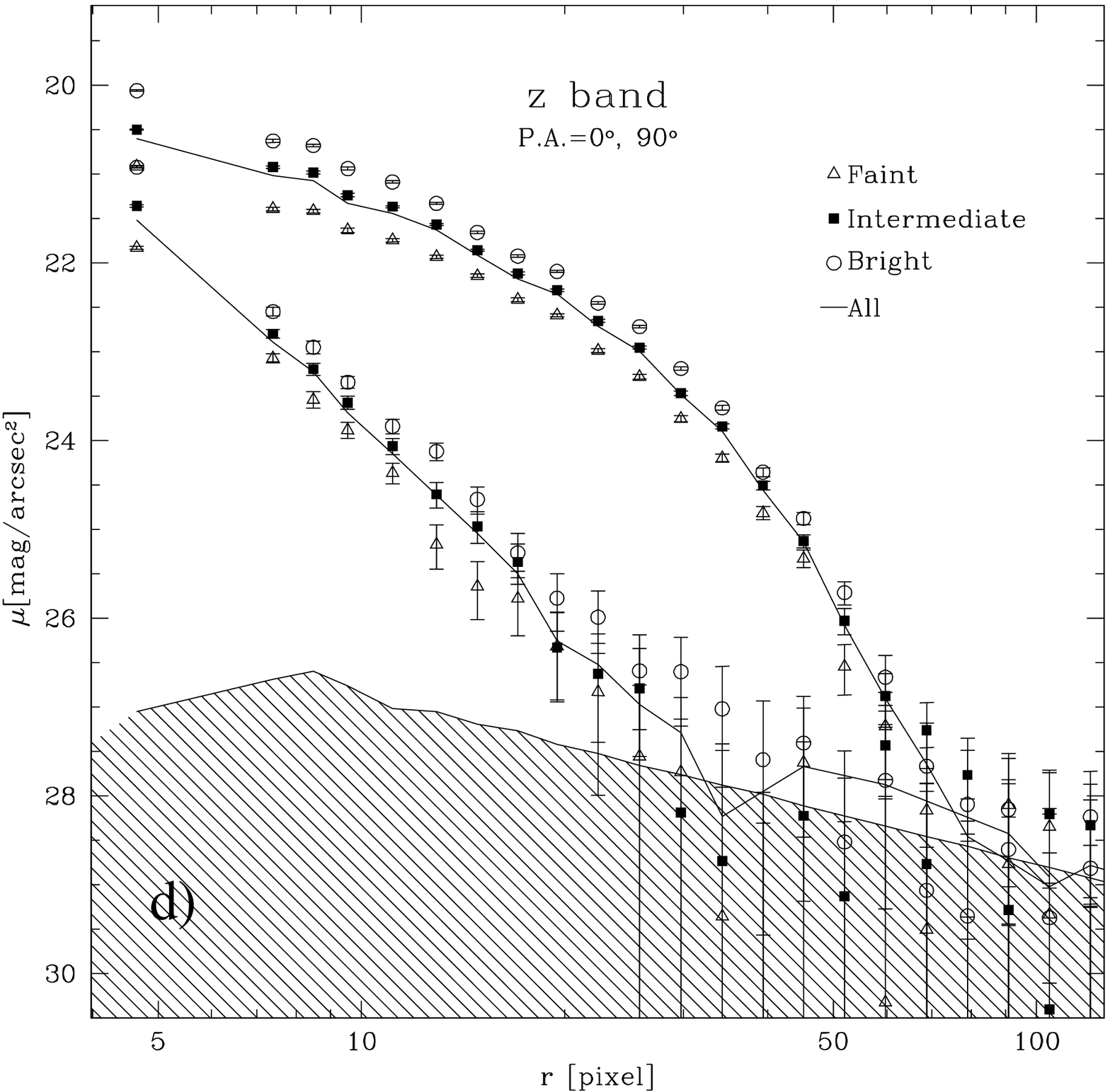}}
\caption{The stacked SB profiles in $g$, $r$, $i$ and $z$ band, at $0^{\mathrm o}$ (upper curves)
and $90^{\mathrm o}$ PA (lower curves), for three luminosity bins: open triangles 
represent the `faint' bin, filled squares the `intermediate' and open circles the `bright'
one. The solid line is the all-sample profile. The shaded area is the rms background fluctuation.
}\label{lumbin}
\end{figure*}

In order to understand the dependence of the halo features on the galaxy luminosity, we 
analyse the stacked images (exponential scale-length rescaling) for three luminosity bins 
in the $i$ band, namely the `bright'
($-22.5 < i_{\mathrm{abs}}+5\log h \leq -20.2$), the `intermediate' ($-20.2 < i_{\mathrm{abs}}+5\log h \leq-19.0$)
and the `faint' one $-19.0 < i_{\mathrm{abs}}+5\log h \leq-16.0$).
The profiles at $0^{\mathrm o}$ and $90^{\mathrm o}$ PA (obtained as in Sec. \ref{profile_sect}) 
are plotted in Fig.
\ref{lumbin}: open triangles for the `faint' bin, filled squares for the `intermediate', and
open circles for the `bright'. The solid lines running through the points are the profiles
as obtained from the complete sample. We notice that the halo SB is correlated with the luminosity
of the galaxy, with the less luminous ones having also fainter halos. A similar correlation with the
total luminosity holds for the disk SB as well, indicating that the relative brightness
of the halo with respect to the disk is roughly constant.
This luminosity dependence is barely observable
in the $g$ band, but becomes increasingly evident at longer wavelengths, with an average offset
of $\sim 0.5~\mathrm {mag~arcsec^{-2}}$ between the `bright' and the `faint' profile in the $i$
band. This also implies that the halos of the bright galaxies are redder than the faint ones by
some $0.1~\mathrm {mag}$ in both $g-r$ and $r-i$ colours.

\section{Discussion}\label{discussion_sec}
The analysis performed in the previous section provides strong evidences for the generalised
presence of a diffuse, low-surface brightness stellar component around disk galaxies. By stacking 
more than 1000 images, we are able to extract reliable photometry at SB level as faint as
$\sim 31~\mathrm{mag~arcsec}^{-2}$ in the $g$, $r$ and $i$ band and $\sim 28~\mathrm{mag~arcsec}^{-2}$
in the $z$ band.\\
It is worth stressing that the statistical estimator adopted in this work to compute the combined
images is extremely robust. The evaluation of the $mode$ for each pixel in the composite image
as $3\times median - 2 \times average$, with the $average$ calculated over the count distribution
after rejecting the 16 per cent percentile tails and the masked pixels, is effective in removing outliers and 
spurious contributions from other sources, and in correcting for the skewness of the distribution,
while reducing the noise of the standard $mode$ estimation.\\
The isophotal contours superposed to the images in Fig. \ref{contourfig} show the
presence of luminous halos whose shape is clearly rounder than the highly flattened central 
disk. The shape of the effective PSF, obtained  from the stacking of 1000 stars selected from 
the same frames and with similar criteria as the galaxies, after applying the same geometrical
transformations, is not consistent with scattered light and extended PSF wings giving
a major contribution to the light detected in the halo. 
Comparisons with the light distribution expected from simple thin+thick disk models, in which we
have accounted for the whole range of galaxy inclinations and for the effective PSF, demonstrate
that no exponential vertical disk component can yield the observed power-law shape of the
SB profiles at $r>4~r_{\mathrm{exp}}$ in any of the four considered bands. Adding a moderately flattened
($c/a\sim 0.6$), power-law ($I\propto r^{-2}$ or $\rho\propto r^{-3}$) halo component to a simple exponential disk, it is 
possible to reproduce the basic features in the observed profiles quite well.\\
It could be argued, that the emission we detect in the outer regions is produced by small bulges.
However, \cite*{macarthur_courteau_holtzmann03} have shown that the bulges of disk-dominated
galaxies, like the ones in our sample, are in general well represented by exponential laws
with $h_{\mathrm{bulge}}/h_{\mathrm{disk}}=0.13\pm0.06$ ($h_{\mathrm{bulge}}$ and
$h_{\mathrm{disk}}$ being the exponential scale-length of the bulge and of the disk respectively).
Assuming an upper limit of $\mu_r=18-19~\mathrm{mag~arcsec^{-2}}$ for the central surface brightness
of the bulge, this implies that bulges contribute no more than $30~\mathrm{mag~arcsec^{-2}}$
beyond 20 pixels. The light we have measured is thus not a simple outward extension of the small
bulges often seen in late-type galaxies.\\
We have shown in Fig. \ref{cfrescale} that the results obtained are independent 
of the scale-length adopted for rescaling the galaxies: this is a strong indication for the absence
of any significant dependence of halo characteristics on the detailed shape of the disk SB profile, even
if it does not necessarily imply complete homology of the disks in our sample.\\
\\
These results are consistent with some known features of the Milky Way halo. Confirming
previous results by \cite{harris76}, \cite{zinn85} derived a halo stellar density
declining as $r^{-3.5}$ from the distribution of the globular clusters. 
Similar results were obtained from the
RR Lyrae distribution by \cite{saha85}, $\rho\propto r^{-3.5}$, by \cite*{preston_etal91}, 
$\rho\propto r^{-3.2\pm 0.1}$, and by \cite{ivezic_etal00}, $\rho\propto r^{-2.7\pm0.2}$. The spatial distribution of Blue Horizontal Branch (BHB)
stars (\citealt{preston_etal91}; \citealt*{kinman_etal94}) is in good good agreement with those estimates,
yielding $\rho\propto r^{-3.5}$ at height $z\gtrsim 5~\mathrm{kpc}\sim~2.5~r_{\mathrm{exp}}$ above the 
disk plane. The SB provided by the MW halo stars is expected to be roughly of the order of 
$30~\mathrm{mag~arcsec}^{-2}$ at $r\sim 8~\mathrm{kpc}\sim 4~r_{\mathrm{exp}}$ in V band 
\citep[][ par. 10.5]{binneymerrifield}, consistently with our measurements in $g$ band.
The analysis by \cite{hartwick87} and by \cite{preston_etal91},
considering RR Lyrae and BHB stars, provides an estimate for moderate flattening of the
spheroidal halo $c/a\sim 0.6$, which is also consistent with the findings in this work.\\
Comparisons with analogous studies of external galaxies are in general difficult. 
Although its
small distance makes M31 the easiest target for observing of the halo population
of a disk galaxy, the prominence of its bulge makes it hard to disentangle the density 
distribution of the halo so that it is perhaps more appropriate to talk about a generic spheroid.
\cite{PvdB94} measured the SB of M31 spheroid to $\mu_V\sim 30~\mathrm{mag~arcsec}^{-2}$ and
concluded that it can be modelled either by a de Vaucouleurs law or, 
in its outer parts, by a power-law $\rho\propto r^{-5}$, which
is much steeper than what we find. However, the globular cluster distribution
follows $\rho\propto r^{-3}$ \citep{racine91} and there is evidence for a shallower
power-law index in the outer parts of the halo from more recent observations 
(Irwin, private communication).\\
Recent observations of the red giant stars of the nearby, late-type spiral
M33 (Ferguson et al. 2003, in preparation), seem to exclude the presence
of a spheroidal component around this galaxy, but its nearly face-on aspect
makes it difficult to draw firm conclusions.\\
Results from more distant galaxies are even more uncertain, because of the overwhelming 
difficulties in going
deeper than $28~\mathrm{mag~arcsec}^{-2}$. After the first claim by \cite{sackett94}
of the detection of halo emission from NGC5907, many discrepant measurements have 
been made by different groups
in bands from the optical to the NIR \citep[see e.g.][]{barnaby_thronson94,lequeux_etal96,rudy_etal97,
james_casali98,lequeux_etal98}. The latest observations by \cite{zheng_etal99} with intermediate-band
filters, and by \cite{yost_etal00} in optical and NIR, together with the RGB
star counts derived by \cite{zepf_etal00} from NICMOS observations, seem to rule
out the presence of a halo, 
favouring instead a luminous ring produced by the tidal disruption of a dwarf companion.\\
However, despite the non-detection of any diffuse component in the Scd galaxy NGC 4244
by \cite{fry_etal99}, probably because of the low sensitivity of their observations
($\mu_R<27.5~\mathrm{mag~arcsec}^{-2}$), many studies during
the last years support the idea of a luminous envelope (thick disk or
halo) surrounding many of disk galaxies. \cite{morrison_etal97} detected
thick disk emission from NGC 891; \cite{abe_etal99} have measured $R$ and $I$ light
excesses with respect to an exponential disk model in the Scd galaxy
IC 5249; \cite{wu_etal02} have observed NGC 4565 at $6600$\AA ~obtaining good
accuracy photometry as faint as $\mu=27.5~\mathrm{mag~arcsec}^{-2}$ and found
a halo component with power-law $r^{-2.3--4.0}$. Similar 
results are found by \cite{rauscher_etal98} in the NIR K band for ESO 240-G11 (power-law
halo $\rho\propto r^{-3.5}$). After observing a sample of 47 extremely flat 
galaxies in $B$, $R$ and $K_s$ down to extremely faint SB, \cite{dalcanton_bernstein02} 
have claimed the ubiquitous presence of red stellar envelopes around
disk galaxies. It is not clear whether these envelopes extend to a spheroidal
halo or are just thick disks, as the authors claim, because
their detection limits are just at the surface brightness  where we start
to see our power-law halo component unambiguously\\
In order to establish the origins of these stellar populations, very precise colour
measurements are needed. Due to the enormous observational difficulties, reliable
colour measurements of the halos around disks are very scarce: excluding the MW,
M31 and M33 for which colour-magnitude diagrams of halo stars and globular clusters
can be obtained, the only optical colours
available to now are the ones derived by \cite{lequeux_etal98} for NGC 5907.
Beside these, \cite{dalcanton_bernstein02} have measured $(B-R)$ and $(R-K_s)$ for
the thick disks in their sample.
The colours we derived in Sec. \ref{colours_sect} are, in fact, very uncertain,
both because of the intrinsic error of the measurements and because of the
practical impossibility to avoid contributions from the disk component. The analysis
in Sec. \ref{colours_sect} showed that scattered light from the disk component
is likely to affect the colours by a few hundredths of a mag, but colour excesses up to
0.2 mag cannot be ruled out.
If we trust these results ($g-r=0.65\pm0.1$, $r-i=0.60\pm0.1$), we 
find that halos are made of stars which are only
marginally ($2 \sigma$) consistent with old, moderately metal-poor stellar populations.
\begin{figure}
\centerline{\includegraphics[width=9truecm]{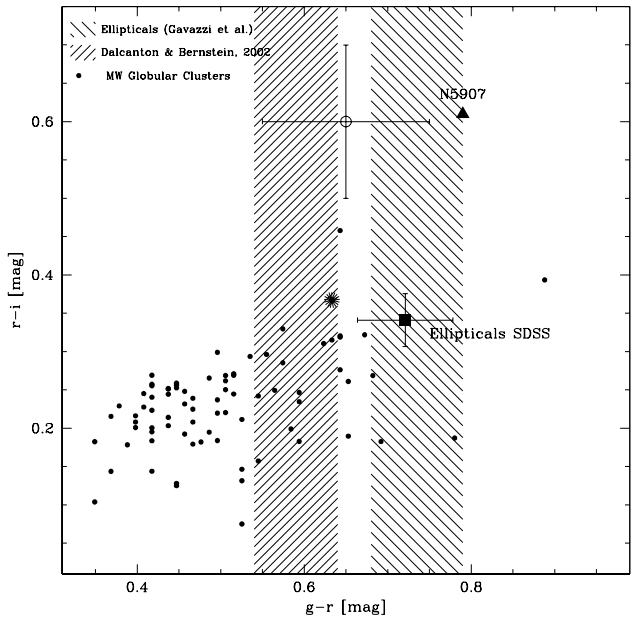}}
\caption{Comparison of the halo colours with other stellar systems in
the $(r-i)$ vs $(g-r)$ diagram. The open circle with error-bars is our measurement,
filled circles are the MW globular clusters from Harris (1996),
the filled square with error-bar
is the average$\pm$rms of the SDSS ellipticals at $z=0.05$ from Bernardi et al. (2003), and
the triangle is the measure of NGC 5907 from Lequeux et al. (1998). The
shaded areas represent the $g-r$ range for the thick disks in 
Dalcanton \& Bernstein (2002) and the interquartile range of the Virgo and
Coma ellipticals from GOLDMiNe (Gavazzi et al. 2003, back hatches).}\label{colour_comp}
\end{figure}
In Fig. \ref{colour_comp} we compare the stacked halo $g-r$, $r-i$ colours with
different data taken from the literature\footnote{Photometric transformations
between different standards are taken from \cite{smith_etal02}.}. 
Our point (open circle with error bars) is almost inconsistent with the colours
of the MW 
globular clusters \citep[filled circles, from][]{MWGC}, being much redder in $r-i$,
and as red as the metal-rich tail of the globular cluster distribution in $g-r$.
We have highlighted 47 Tucanae in the plot (starred dot), because its integrated
colours are very similar to M31's halo. Its $g-r$ is about the
same as our measurements.
We note that $g-r$ is also consistent with the blue end of the elliptical sequence
as derived from the SDSS at the median redshift of the sample $z=0.05$ 
\citep[filled square, ][]{bernardi03_IV}, 
and from the observations in Virgo and Coma
by Gavazzi et al. in B and V, as given in GOLDMiNe \citep{goldmine}, whose
interquartile range is represented by the back-shaded area.
Reasonable agreement is found with the optical colours $B-R$ derived by
\cite{dalcanton_bernstein02} for their thick disks (whose range is represented 
by the shaded area in Fig. \ref{colour_comp}).
Reconciling the measured $r-i$ colour with any known stellar population is
almost impossible, even if we allow for an extreme 0.1 mag reddening caused 
by the PSF, as discussed in Sec. \ref{colours_sect}.
It is however interesting to note that the colours derived for
NGC 5907 by \cite{lequeux_etal98} are consistent with ours, but nonetheless 
troublesome. We exclude significant dust reddening for two
reasons. First, the unusually red colour in our data is $r-i$, despite the
fact that dust mostly affects optical/blue bands. Second, in the $r-i$ profile
there is quite strong evidence for a red gradient toward the outer parts,
whereas it is known that the dust is concentrated in a thin layer in the disk.
Thus we conclude that
there is evidence for the halos being made of extremely red stellar 
populations. This is likely to be primarily due to an old age, but other
effects, such as high metallicity or `exotic' low-mass dominated IMF, 
would be required in order to explain the unusually high $r-i$.\\

Our red colours seem to exclude the possibility that the majority
of the halos we observe around disk galaxies are made of metal-poor
stars, or that they result from the integrated light of  
globular cluster populations. 
The correlation between the disk and halo SB suggests
a link between the two components.
On the other hand, the high latitude extension of the emission
rules out `disk heating' as an effective formation mechanism.
An accretion (or `cannibalism') scenario, in which the halo is built up by
capture and disruption of spheroidal satellites, presents
many advantages in explaining the observations. In this scenario dwarf spheroidal
galaxies, made of old, metal enriched stars, are tidally disrupted by the
gravitational field of the central galaxy and their stars are spread to fill 
the phase space almost isotropically in a few dynamical times. Thus both the
spheroidal shapes and the extremely red colours of the halos could be explained,
at least qualitatively. As already mentioned, recent observations in the halo 
of MW and M31, of the Sagittarius stream in the MW,
and of the ring in the halo of NGC 5907
support the idea that this mechanism has been working till very
recent times and is not uncommon among the disk galaxies.\\

Our analysis in different luminosity bins shows that the
halo luminosity, at least on average, is proportional to disk luminosity.
Prominent halos are in the more luminous galaxies. The decrease of the
average surface brightness at lower luminosity affects mostly the redder
pass-bands and is reflection of the well known correlation of disk
surface brightness with disk luminosity in the red bands \citep[see e.g.][]{shen_etal03}.
The light we see does not appear to be a straightforward extension of the bulge,
since the power-law shapes of 
the profiles, irrespective of luminosity, are not consistent 
with classical de Vaucouleurs or exponential bulges. We can interpret the
halo-disk luminosity relation we find here in the hierarchical picture, in which
more luminous galaxies sit in more massive DM halos, with a larger number
of merging sub-halos contributing to the stellar halo luminosity.
However, deeper and more detailed observations, along with more reliable 
theoretical predictions for the number, stellar mass and metal content of the
accreted satellites, are needed in order to confirm this hypothesis.\\

\section{Summary and conclusions}\label{conclusions}
By stacking a large number ($>1000$) of edge-on disk galaxies imaged in the SDSS
we have been able to detect a diffuse, spheroidal, low-surface brightness 
component around the disk. This detection is significant 
in the $g$, $r$, $i$ and $z$ bands, and cannot be ascribed to any obvious 
instrumental artifact (e.g. scattered light or PSF). Given the statistical estimator
we adopt for combining the images, our result indicates that a substantial
fraction of the stacked galaxies must share the observed halo characteristics,
even if we cannot exclude the possibility that a number of disk galaxies actually have
no halo at all.
The halo can be described by a power-law projected density profile
$I\propto r^{-\alpha}$, with $\alpha\sim2$ nearly irrespective of the band.\\
The colour measurements provide inconclusive and troublesome results, but there is
a clear indication for extremely red colours. $g-r$ is consistent
with old, moderately metal-poor stellar populations, such as the more metal-rich
MW's globular clusters, 47 Tucanae, the halo of M31 and the most metal-poor 
ellipticals. $r-i$ is (at $2~\sigma$) 0.2 mag redder than the reddest known stellar populations
in globular clusters and elliptical galaxies
and it is difficult to reconcile with any theoretical models, even allowing for
\emph{ad hoc} modified IMF's dominated by low-mass stars and high metallicity.
The data also suggest a correlation between the luminosity of the halo and the
total luminosity of the galaxy.\\
The results presented in this work are far from being conclusive, but nevertheless
they are consistent with the idea that a large fraction of disk galaxies
are surrounded by a luminous halo. The colours, although affected
by large uncertainties, hint at old, but not particularly metal-poor stellar 
populations, thus supporting a scenario in which the halos are mostly contributed
by stars stripped from accreted or merged companions, in which the chemical
evolution was already advanced.\\
Deeper, individual observations of a large sample of nearby galaxies will
be required, however, in order to assess the validity of this scenario by 
quantifying not only the average halo, but also the whole distribution
of individual halo  parameters.\\

\thanks{We wish to thank the referees James Lequeux and Francoise Combes for
helpful advice.\\
S.Z. wishes to thank Annette Ferguson, St\'ephane Charlot, Amina Helmi
and Tim McKay for the useful discussions.}
\\

Funding for the creation and distribution of the SDSS Archive has been provided 
by the Alfred P. Sloan Foundation, the Participating Institutions, the National 
Aeronautics and Space Administration, the National Science Foundation, the U.S. 
Department of Energy, the Japanese Monbukagakusho, and the Max Planck Society. 
The SDSS Web site is http://www.sdss.org/.\\
The SDSS is managed by the Astrophysical Research Consortium (ARC) for the 
Participating Institutions. The Participating Institutions are The University 
of Chicago, Fermilab, the Institute for Advanced Study, the Japan Participation 
Group, The Johns Hopkins University, Los Alamos National Laboratory, the 
Max-Planck-Institute for Astronomy (MPIA), the Max-Planck-Institute for Astrophysics (MPA), 
New Mexico State University, University of Pittsburgh, Princeton University, 
the United States Naval Observatory, and the University of Washington.\\

\label{lastpage}

\begin{thebibliography}{} 
\bibitem[\protect\citeauthoryear{{Abazajian} et~al.}{{Abazajian} et~al.}{2003}]{DR1} {Abazajian} K., {et~al.} 2003, preprint(astro-ph/0305492) 
\bibitem[\protect\citeauthoryear{{Abe} et~al.}{{Abe} et~al.}{1999}]{abe_etal99} {Abe} F., {et~al.} 1999, \aj, 118, 261 
\bibitem[\protect\citeauthoryear{{Barnaby} \& {Thronson}}{{Barnaby} \& {Thronson}}{1994}]{barnaby_thronson94} {Barnaby} D., {Thronson} H.~A., 1994, \aj, 107, 1717 
\bibitem[\protect\citeauthoryear{{Baugh}, {Cole} \& {Frenk}}{{Baugh} et~al.}{1996}]{baugh_etal96} {Baugh} C.~M., {Cole} S., {Frenk} C.~S., 1996, \mnras, 283, 1361 
\bibitem[\protect\citeauthoryear{{Benson}, {Baugh}, {Cole}, {Frenk} \& {Lacey}}{{Benson} et~al.}{2000}]{benson_etal00} {Benson} A.~J., {Baugh} C.~M., {Cole} S., {Frenk} C.~S., {Lacey} C.~G., 2000, \mnras, 316, 107 
\bibitem[\protect\citeauthoryear{{Bernardi} et~al.}{{Bernardi} et~al.}{2003}]{bernardi03_IV} {Bernardi} M., {et~al.} 2003, \aj, 125, 1882 
\bibitem[\protect\citeauthoryear{{Bertin} \& {Arnouts}}{{Bertin} \& {Arnouts}}{1996}]{sex} {Bertin} E., {Arnouts} S., 1996, \aaps, 117, 393 
\bibitem[\protect\citeauthoryear{{Binney} \& {Merrifield}}{{Binney} \& {Merrifield}}{1998}]{binneymerrifield} {Binney} J., {Merrifield} M., 1998, {Galactic astronomy}. Princeton University Press,~ Princeton, NJ 
\bibitem[\protect\citeauthoryear{{Blanton} et~al.}{{Blanton} et~al.}{2002}]{LSS} {Blanton} M.~R., {et~al.} 2002, preprint(astro-ph/0210215) 
\bibitem[\protect\citeauthoryear{{Blanton}, {Lin}, {Lupton}, {Maley}, {Young}, {Zehavi} \& {Loveday}}{{Blanton} et~al.}{2003}]{blanton_etal03} {Blanton} M.~R., {Lin} H., {Lupton} R.~H., {Maley} F.~M., {Young} N., {Zehavi} I., {Loveday} J., 2003, \aj, 125, 2276 
\bibitem[\protect\citeauthoryear{{Dalcanton} \& {Bernstein}}{{Dalcanton} \& {Bernstein}}{2002}]{dalcanton_bernstein02} {Dalcanton} J.~J., {Bernstein} R.~A., 2002, \aj, 124, 1328 
\bibitem[\protect\citeauthoryear{{Elson}, {Fall} \& {Freeman}}{{Elson} et~al.}{1987}]{halo_powerlaw} {Elson} R.~A.~W., {Fall} S.~M., {Freeman} K.~C., 1987, \apj, 323, 54 
\bibitem[\protect\citeauthoryear{{Ferguson}, {Irwin}, {Ibata}, {Lewis} \& {Tanvir}}{{Ferguson} et~al.}{2002}]{ferguson_etal02} {Ferguson} A.~M.~N., {Irwin} M.~J., {Ibata} R.~A., {Lewis} G.~F., {Tanvir} N.~R., 2002, \aj, 124, 1452 
\bibitem[\protect\citeauthoryear{{Fry}, {Morrison}, {Harding} \& {Boroson}}{{Fry} et~al.}{1999}]{fry_etal99} {Fry} A.~M., {Morrison} H.~L., {Harding} P., {Boroson} T.~A., 1999, \aj, 118, 1209 
\bibitem[\protect\citeauthoryear{{Fukugita}, {Ichikawa}, {Gunn}, {Doi}, {Shimasaku} \& {Schneider}}{{Fukugita} et~al.}{1996}]{fukugita_etal96} {Fukugita} M., {Ichikawa} T., {Gunn} J.~E., {Doi} M., {Shimasaku} K., {Schneider} D.~P., 1996, \aj, 111, 1748 
\bibitem[\protect\citeauthoryear{{Gavazzi}, {Boselli}, {Donati}, {Franzetti} \& {Scodeggio}}{{Gavazzi} et~al.}{2003}]{goldmine} {Gavazzi} G., {Boselli} A., {Donati} A., {Franzetti} P., {Scodeggio} M., 2003, \aap, 400, 451 
\bibitem[\protect\citeauthoryear{{Gunn} et~al.}{{Gunn} et~al.}{1998}]{gunn_etal98} {Gunn} J.~E., {et~al.} 1998, \aj, 116, 3040 
\bibitem[\protect\citeauthoryear{{Harris}}{{Harris}}{1976}]{harris76} {Harris} W.~E., 1976, \aj, 81, 1095 
\bibitem[\protect\citeauthoryear{{Harris}}{{Harris}}{1996}]{MWGC} {Harris} W.~E., 1996, \aj, 112, 1487 
\bibitem[\protect\citeauthoryear{{Hartwick}}{{Hartwick}}{1987}]{hartwick87} {Hartwick} F.~D.~A., 1987, in NATO ASIC Proc. 207: The Galaxy {The structure of the Galactic halo}. pp 281--290 
\bibitem[\protect\citeauthoryear{{Helmi}, {White}, {de Zeeuw} \& {Zhao}}{{Helmi} et~al.}{1999}]{helmi_etal99} {Helmi} A., {White} S.~D.~M., {de Zeeuw} P.~T., {Zhao} H., 1999, \nat, 402, 53 
\bibitem[\protect\citeauthoryear{{Hogg}, {Finkbeiner}, {Schlegel} \& {Gunn}}{{Hogg} et~al.}{2001}]{hogg_etal01} {Hogg} D.~W., {Finkbeiner} D.~P., {Schlegel} D.~J., {Gunn} J.~E., 2001, \aj, 122, 2129 
\bibitem[\protect\citeauthoryear{{Ibata}, {Gilmore} \& {Irwin}}{{Ibata} et~al.}{1994}]{ibata_etal94} {Ibata} R.~A., {Gilmore} G., {Irwin} M.~J., 1994, \nat, 370, 194 
\bibitem[\protect\citeauthoryear{{Ibata}, {Irwin}, {Lewis}, {Ferguson} \& {Tanvir}}{{Ibata} et~al.}{2003}]{ibata_etal03} {Ibata} R.~A., {Irwin} M.~J., {Lewis} G.~F., {Ferguson} A.~M.~N., {Tanvir} N., 2003, \mnras, 340, L21 
\bibitem[\protect\citeauthoryear{{Ivezi{\' c}} et~al.}{{Ivezi{\' c}} et~al.}{2000}]{ivezic_etal00} {Ivezi{\' c}} {\v Z}., {et~al.} 2000, \aj, 120, 963 
\bibitem[\protect\citeauthoryear{{James} \& {Casali}}{{James} \& {Casali}}{1998}]{james_casali98} {James} P.~A., {Casali} M.~M., 1998, \mnras, 301, 280 
\bibitem[\protect\citeauthoryear{{Kauffmann}, {Colberg}, {Diaferio} \& {White}}{{Kauffmann} et~al.}{1999}]{kauffmann_etal99} {Kauffmann} G., {Colberg} J.~M., {Diaferio} A., {White} S.~D.~M., 1999, \mnras, 303, 188 
\bibitem[\protect\citeauthoryear{{Kauffmann}, {Nusser} \& {Steinmetz}}{{Kauffmann} et~al.}{1997}]{kauffmann_etal97} {Kauffmann} G., {Nusser} A., {Steinmetz} M., 1997, \mnras, 286, 795 
\bibitem[\protect\citeauthoryear{{Kauffmann}, {White} \& {Guiderdoni}}{{Kauffmann} et~al.}{1993}]{kauffmann_etal93} {Kauffmann} G., {White} S.~D.~M., {Guiderdoni} B., 1993, \mnras, 264, 201 
\bibitem[\protect\citeauthoryear{{Kinman}, {Suntzeff} \& {Kraft}}{{Kinman} et~al.}{1994}]{kinman_etal94} {Kinman} T.~D., {Suntzeff} N.~B., {Kraft} R.~P., 1994, \aj, 108, 1722 
\bibitem[\protect\citeauthoryear{{Kregel}, {van der Kruit} \& {de Grijs}}{{Kregel} et~al.}{2002}]{kregel_etal02} {Kregel} M., {van der Kruit} P.~C., {de Grijs} R., 2002, \mnras, 334, 646 
\bibitem[\protect\citeauthoryear{{Lequeux}, {Combes}, {Dantel-Fort}, {Cuillandre}, {Fort} \& {Mellier}}{{Lequeux} et~al.}{1998}]{lequeux_etal98} {Lequeux} J., {Combes} F., {Dantel-Fort} M., {Cuillandre} J.-C., {Fort} B., {Mellier} Y., 1998, \aap, 334, L9 
\bibitem[\protect\citeauthoryear{{Lequeux}, {Fort}, {Dantel-Fort}, {Cuillandre} \& {Mellier}}{{Lequeux} et~al.}{1996}]{lequeux_etal96} {Lequeux} J., {Fort} B., {Dantel-Fort} M., {Cuillandre} J.-C., {Mellier} Y., 1996, \aap, 312, L1 
\bibitem[\protect\citeauthoryear{{Lupton}, {Gunn}, {Ivezi{\' c}}, {Knapp}, {Kent} \& {Yasuda}}{{Lupton} et~al.}{2001}]{lupton_etal01} {Lupton} R.~H., {Gunn} J.~E., {Ivezi{\' c}} Z., {Knapp} G.~R., {Kent} S., {Yasuda} N., 2001, in ASP Conf. Ser. 238: Astronomical Data Analysis Software and Systems X {The SDSS Imaging Pipelines}. pp 269--+ 
\bibitem[\protect\citeauthoryear{{MacArthur}, {Courteau} \& {Holtzman}}{{MacArthur} et~al.}{2003}]{macarthur_courteau_holtzmann03} {MacArthur} L.~A., {Courteau} S., {Holtzman} J.~A., 2003, \apj, 582, 689 
\bibitem[\protect\citeauthoryear{{Majewski}}{{Majewski}}{1993}]{majewski93} {Majewski} S.~R., 1993, \araa, 31, 575 
\bibitem[\protect\citeauthoryear{{Morrison}, {Miller}, {Harding}, {Stinebring} \& {Boroson}}{{Morrison} et~al.}{1997}]{morrison_etal97} {Morrison} H.~L., {Miller} E.~D., {Harding} P., {Stinebring} D.~R., {Boroson} T.~A., 1997, \aj, 113, 2061 
\bibitem[\protect\citeauthoryear{{Navarro} \& {Steinmetz}}{{Navarro} \& {Steinmetz}}{2000}]{navarro_steinmetz00} {Navarro} J.~F., {Steinmetz} M., 2000, \apj, 538, 477 
\bibitem[\protect\citeauthoryear{{Navarro} \& {White}}{{Navarro} \& {White}}{1994}]{navarro_white94} {Navarro} J.~F., {White} S.~D.~M., 1994, \mnras, 267, 401 
\bibitem[\protect\citeauthoryear{{Odenkirchen}, {Grebel}, {Dehnen}, {Rix} \& {Cudworth}}{{Odenkirchen} et~al.}{2002}]{odenkirchen_etal02} {Odenkirchen} M., {Grebel} E.~K., {Dehnen} W., {Rix} H., {Cudworth} K.~M., 2002, \aj, 124, 1497 
\bibitem[\protect\citeauthoryear{{Petrosian}}{{Petrosian}}{1976}]{petrosian76} {Petrosian} V., 1976, \apjl, 209, L1 
\bibitem[\protect\citeauthoryear{{Pier}, {Munn}, {Hindsley}, {Hennessy}, {Kent}, {Lupton} \& {Ivezi{\' c}}}{{Pier} et~al.}{2003}]{pier_etal03} {Pier} J.~R., {Munn} J.~A., {Hindsley} R.~B., {Hennessy} G.~S., {Kent} S.~M., {Lupton} R.~H., {Ivezi{\' c}} {\v Z}., 2003, \aj, 125, 1559 
\bibitem[\protect\citeauthoryear{{Preston}, {Shectman} \& {Beers}}{{Preston} et~al.}{1991}]{preston_etal91} {Preston} G.~W., {Shectman} S.~A., {Beers} T.~C., 1991, \apj, 375, 121 
\bibitem[\protect\citeauthoryear{{Pritchet} \& {van den Bergh}}{{Pritchet} \& {van den Bergh}}{1994}]{PvdB94} {Pritchet} C.~J., {van den Bergh} S., 1994, \aj, 107, 1730 
\bibitem[\protect\citeauthoryear{{Racine}}{{Racine}}{1991}]{racine91} {Racine} R., 1991, \aj, 101, 865 
\bibitem[\protect\citeauthoryear{{Rauscher}, {Lloyd}, {Barnaby}, {Harper}, {Hereld}, {Loewenstein}, {Severson} \& {Mrozek}}{{Rauscher} et~al.}{1998}]{rauscher_etal98} {Rauscher} B.~J., {Lloyd} J.~P., {Barnaby} D., {Harper} D.~A., {Hereld} M., {Loewenstein} R.~F., {Severson} S.~A., {Mrozek} F., 1998, \apj, 506, 116 
\bibitem[\protect\citeauthoryear{{Rudy}, {Woodward}, {Hodge}, {Fairfield} \& {Harker}}{{Rudy} et~al.}{1997}]{rudy_etal97} {Rudy} R.~J., {Woodward} C.~E., {Hodge} T., {Fairfield} S.~W., {Harker} D.~E., 1997, \nat, 387, 159 
\bibitem[\protect\citeauthoryear{{Sackett}, {Morrison}, {Harding} \& {Boroson}}{{Sackett} et~al.}{1994}]{sackett94} {Sackett} P.~D., {Morrison} H.~L., {Harding} P., {Boroson} T.~A., 1994, \nat, 370, 441 
\bibitem[\protect\citeauthoryear{{Saha}}{{Saha}}{1985}]{saha85} {Saha} A., 1985, \apj, 289, 310 
\bibitem[\protect\citeauthoryear{{Scannapieco} \& {Tissera}}{{Scannapieco} \& {Tissera}}{2003}]{scannapieco_tissera03} {Scannapieco} C., {Tissera} P.~B., 2003, \mnras, 338, 880 
\bibitem[\protect\citeauthoryear{{Schlegel}, {Finkbeiner} \& {Davis}}{{Schlegel} et~al.}{1998}]{schlegel_dust} {Schlegel} D.~J., {Finkbeiner} D.~P., {Davis} M., 1998, \apj, 500, 525 
\bibitem[\protect\citeauthoryear{{Shen}, {Mo}, {White}, {Blanton}, {Kauffmann}, {Voges}, {Brinkmann} \& {Csabai}}{{Shen} et~al.}{2003}]{shen_etal03} {Shen} S., {Mo} H.~J., {White} S.~D.~M., {Blanton} M.~R., {Kauffmann} G., {Voges} W., {Brinkmann} J., {Csabai} I., 2003, preprint(astro-ph/0301527)
\bibitem[\protect\citeauthoryear{{Smith} et~al.}{{Smith} et~al.}{2002}]{smith_etal02} {Smith} J.~A., {et~al.} 2002, \aj, 123, 2121 
\bibitem[\protect\citeauthoryear{{Sommer-Larsen}, {Gelato} \& {Vedel}}{{Sommer-Larsen} et~al.}{1999}]{sommerlarsen_etal99} {Sommer-Larsen} J., {Gelato} S., {Vedel} H., 1999, \apj, 519, 501 
\bibitem[\protect\citeauthoryear{{Somerville} \& {Primack}}{{Somerville} \& {Primack}}{1999}]{somerville_primack99} {Somerville} R.~S., {Primack} J.~R., 1999, \mnras, 310, 1087 
\bibitem[\protect\citeauthoryear{{Springel}, {White}, {Tormen} \& {Kauffmann}}{{Springel} et~al.}{2001}]{springel_etal01} {Springel} V., {White} S.~D.~M., {Tormen} G., {Kauffmann} G., 2001, \mnras, 328, 726 
\bibitem[\protect\citeauthoryear{{Stoughton} et~al.}{{Stoughton} et~al.}{2002}]{EDR} {Stoughton} C., {et~al.} 2002, \aj, 123, 485 
\bibitem[\protect\citeauthoryear{{Strauss} et~al.}{{Strauss} et~al.}{2002}]{strauss_etal02} {Strauss} M.~A., {et~al.} 2002, \aj, 124, 1810 
\bibitem[\protect\citeauthoryear{{van der Kruit}}{{van der Kruit}}{2001}]{vanderkruit01} {van der Kruit} P.~C., 2001, in ASP Conf. Ser. 230: Galaxy Disks and Disk Galaxies {Truncations in Stellar Disks}. pp 119--126 
\bibitem[\protect\citeauthoryear{{Wu} et~al.}{{Wu} et~al.}{2002}]{wu_etal02} {Wu} H., {et~al.} 2002, \aj, 123, 1364 
\bibitem[\protect\citeauthoryear{{Yanny}, {Newberg}, {Grebel}, {Kent}, {Odenkirchen}, {Rockosi}, {Schlegel}, {Subbarao}, {Brinkmann}, {Fukugita}, {Ivezic}, {Lamb}, {Schneider} \& {York}}{{Yanny} et~al.}{2003}]{yanny_etal03} {Yanny} B., {Newberg} H.~J., {Grebel} E.~K., {Kent} S., {Odenkirchen} M., {Rockosi} C.~M., {Schlegel} D., {Subbarao} M., {Brinkmann} J., {Fukugita} M., {Ivezic} Z., {Lamb} D.~Q., {Schneider} D.~P., {York} D.~G., 2003, \apj, 588, 824 
\bibitem[\protect\citeauthoryear{{York} et~al.}{{York} et~al.}{2000}]{SDSS} {York} D.~G., {et~al.} 2000, \aj, 120, 1579 
\bibitem[\protect\citeauthoryear{{Yost}, {Bock}, {Kawada}, {Lange}, {Matsumoto}, {Uemizu}, {Watabe} \& {Wada}}{{Yost} et~al.}{2000}]{yost_etal00} {Yost} S.~A., {Bock} J.~J., {Kawada} M., {Lange} A.~E., {Matsumoto} T., {Uemizu} K., {Watabe} T., {Wada} T., 2000, \apj, 535, 644 
\bibitem[\protect\citeauthoryear{{Zepf}, {Liu}, {Marleau}, {Sackett} \& {Graham}}{{Zepf} et~al.}{2000}]{zepf_etal00} {Zepf} S.~E., {Liu} M.~C., {Marleau} F.~R., {Sackett} P.~D., {Graham} J.~R., 2000, \aj, 119, 1701 
\bibitem[\protect\citeauthoryear{{Zheng} et~al.}{{Zheng} et~al.}{1999}]{zheng_etal99} {Zheng} Z., {et~al.} 1999, \aj, 117, 2757 
\bibitem[\protect\citeauthoryear{{Zinn}}{{Zinn}}{1985}]{zinn85} {Zinn} R., 1985, \apj, 293, 424 
\end{thebibliography}
\end{document}